\documentclass[
    prd,
    twocolumn,
    superscriptaddress,
    floatfix,
    amsmath,
    amssymb,
    nofootinbib,
    longbibliography
]{revtex4-2}

\usepackage[normalem]{ulem}
\usepackage{graphicx} 
\usepackage{amsthm}
\usepackage{amsmath}
\usepackage{amssymb}
\usepackage[scr=rsfs]{mathalpha}
\usepackage{mathtools}
\usepackage{color}
\usepackage{bm}
\usepackage{hyperref}
\usepackage{cancel}
\usepackage{comment}
\usepackage{tcolorbox}
\usepackage{changepage}
\usepackage{tikz}
\usepackage{circuitikz}
\usepackage{physics}
\newcommand{\mf}{\mathsf} 
\renewcommand{\rm}{\mathrm} 
\newcommand{\ii}{\mathrm{i}}
\newcommand{\proj}[2]{\left| {#1} \right\rangle\!\left\langle {#2} \right|}

\renewcommand{\H}{\mathcal{H}}
\newcommand{\Hil}{\mathscr{H}}

\renewcommand{\L}{\mathcal{L}}

\newcommand{\tc}[1]{\textsc{#1}}

\renewcommand{\paragraph}[1]{\subsection{#1}}

\def\openone{\leavevmode\hbox{\small$1$\normalsize\kern-.33em$1$}}

\newtheorem{definition}{Definition}[section]

\begin{document}

\title{Causality in relativistic quantum interactions without mediators}


\author{Eirini Telali}
\email{etelali@perimeterinstitute.ca}
\affiliation{Perimeter Institute for Theoretical Physics, Waterloo, Ontario, N2L 2Y5, Canada}
\affiliation{Department of Physics and Astronomy, University of Waterloo, Waterloo, Ontario, N2L 3G1, Canada}

\author{T. Rick Perche}
\email{trickperche@perimeterinstitute.ca}
\affiliation{Perimeter Institute for Theoretical Physics, Waterloo, Ontario, N2L 2Y5, Canada}
\affiliation{Department of Applied Mathematics, University of Waterloo, Waterloo, Ontario, N2L 3G1, Canada}
\affiliation{Institute for Quantum Computing, University of Waterloo, Waterloo, Ontario, N2L 3G1, Canada}

\author{Eduardo Mart\'in-Mart\'inez}
\email{emartinmartinez@uwaterloo.ca}
\affiliation{Perimeter Institute for Theoretical Physics, Waterloo, Ontario, N2L 2Y5, Canada}
\affiliation{Department of Applied Mathematics, University of Waterloo, Waterloo, Ontario, N2L 3G1, Canada}
\affiliation{Institute for Quantum Computing, University of Waterloo, Waterloo, Ontario, N2L 3G1, Canada}

\begin{abstract}

    We analyse the interaction between two quantum systems in spacetime and we compare two possible models to describe it: 1) a fully quantum field theoretical (QFT)  description of the coupling of two quantum systems mediated by a quantum field and 2) a quantum-controlled model (qc-model), which is an effectively relativistic direct-coupling  in which the interaction of two quantum systems is not mediated by a field with local quantum degrees of freedom. We show that while there are regimes where the qc-model can approximate QFT arbitrarily well, it can suffer from retrocausal effects. We discuss in what regimes those retrocausal predictions of the qc-model are non-negligible and whether they can be used to argue that gravity induced entanglement experiments can reveal genuinely quantum aspects of the gravitational interaction or not.
\end{abstract}

\maketitle

\section{Introduction}


Relativistic quantum field theories (QFTs) provide our best consistent foundational framework for describing physical interactions between quantum systems.   
However, in many situations and for practical considerations, one does not require a full description of the quantum degrees of freedom of the mediating fields. Examples of these setups are the spin-coupling in electron pair resonance~\cite{esrReviewDavid,esrspincavity}, nuclear magnetic resonance~\cite{BlochT1T2}, among others. In general, effective non-relativistic direct-coupling theories are often used everywhere.
It is rarely the case that the full power of QFT is required to model low-energy physics, quantum information tasks or, in general, to describe non-relativistic experiments.



A perhaps less-known fact is that direct-coupling models are not entirely restricted to non-relativistic regimes. 
Indeed, one can formulate interactions between two quantum systems in spacetime without the interaction propagating instantaneously using the so-called quantum-controlled (qc) model (see, for example, ~\cite{quantClass,ourBMV})). 
In essence, the only relevant degrees of freedom in a quantum-controlled description are associated with the quantum sources, and relativity is incorporated through retarded propagators of a classical mediating field theory in the direct coupling. This effective model has been shown to accurately approximate the predictions of the QFT description of interactions that are prolonged in time, both in light-matter like setups and gravitational scenarios such as gravity mediated entanglement experiments~\cite{quantClass,ourBMV}.

Quantifying the regimes of applicability of the qc-model is important not only to simplify the description of experiments that do not require a full QFT approach. 
It is also a tool for identifying what physical processes  explicitly rely on the quantum degrees of freedom of mediating fields: if the qc-model is enough to describe the experiment, the quantum degrees of freedom of the mediators do not play any significant role in them. 
Understanding the role of local quantum degrees of freedom is particularly relevant in the context of gravity mediated entanglement experiments (GME)~\cite{B,MV}, where it has been proposed that gravitationally induced entanglement can be used to witness quantum degrees of freedom of the gravitational field~\cite{CHRISTODOULOU201964,Christodoulou_2023,PhysRevLett.130.100202}.
However, in~\cite{ourBMV} it was shown that within the regimes originally proposed in~\cite{B,MV}, GME experiments can be accurately described in terms of a qc-model, showcasing that quantum degrees of freedom of the gravitational field are not directly accessed by the experiment\footnote{Notice that there are other sources of criticism towards the ability of some GME experiments to reveal the quantum nature of gravity. They are not related to the points we want to make in this paper buf for completeness see~\cite{CharisHu,treta,tretaReply,tretaRereply,Anastopoulos2021,Julen}.}.

Although qc-models can accurately describe many physical setups, the direct coupling between two systems in different spacetime locations (even if the coupling is retardedly propagated) may lead to some degree of retrocausal effects~\cite{FlaminiaQuantumGravity}. 
These effects do not fully prevent the applicability of qc-models, but certainly provide bounds to their regime of validity: one could not argue that a qc-model can describe a setup if the retrocausal predictions are of the same order as the resolution of the experiments. Our goal in this manuscript is to study in what scenarios retrocausality manifests in the qc-model, quantify the retrocausal effects, and discuss the consequences of this to the applicability of the model, with a special attention to GME experiments.

To classify retrocausal effects, we explicitly define conditions that must be fulfilled by general causal interactions in spacetime. Applying these definitions to qc-interactions, we can split the predictions of the model into causal and retrocausal contributions. This allows for a quantitative analysis of retrocausal effects, naturally providing a limit to the regime of validity of quantum-controlled effective models. 

In this study,  we will show that retrocausal effects become arbitrarily negligible in the limit where both systems maintain causal contact for long times relative to their spatial separation. This is relevant because these are precisely the the setups in the current implementations of GME experimental proposals.

This manuscript is organized as follows. In Section~\ref{sec:QCQFT} we review the qc-model for interactions between quantum systems and compare it interactions mediated by fully featured quantum fields. In Section~\ref{sec:Retrocausal} we discuss retrocausality in interactions between two quantum systems in spacetime, presenting definitions of causal and non-retrocausal interactions and establish a relationship between causality violations in the qc-model and the regimes where it approximates a QFT description. By studying explicit examples, we quantify retrocausal effects in relativistic quantum-controlled interactions in Section~\ref{sec:quantify}, discussing the implications for gravity mediated entanglement experiments. The conclusions of our work can be found in Section~\ref{sec:conclusions}.

\section{Quantum Controlled fields as a Quantum Field Theory limit}\label{sec:QCQFT}


 
A particular case of an effective interaction between two quantum systems, where the interaction has relativistic elements but is not entirely quantum is the so-called \textit{quantum-controlled field} (qc-field) model~\cite{quantClass,ourBMV}. 
This model replaces the quantum degrees of freedom of a mediating field by a direct coupling between quantum systems taking into account the relativistic propagation of the interaction. 
In this section we will review the formalism of qc-fields and discuss the limit where a quantum field theoretic model gives rise to these effective models.

\subsection{Quantum Controlled fields} \label{subsec:QuantumControlled}

We start with a brief review of the direct coupling of two systems within the qc-field formalism. 
We motivate the model by first describing the interaction between two classical systems through a relativistic classical field. 
This description is then `upgraded' to qc-field model by letting the interacting systems be quantum, which gives rise to a direct interaction between the two systems---with no intermediate field degrees of freedom---that is dictated by the appropriate relativistic retarded potentials.

\subsubsection{Classical sources interacting via classical fields}\label{sub:ClassicalModel}

Assume two classical systems A and B in Minkowski spacetime, $\mathcal{M}$, that interact via a classical field $\phi^{(a)\!}$, where $(a)\equiv (\mu,\nu,\dots)$ is an arbitrary set of Lorentz indices. 
The two systems couple to the field through currents $j^{\tc{a}}_{(a)\!}$, $j^{\tc{b}}_{(a)\!}$. 
The total system is described by the action:
\begin{align}\label{eq:LagrangianClassical}
S &= \!\int \!\dd V \! 
 \left( \L_{\phi} + \L_{\tc{a}} +\L_{\tc{b}} \right.\nonumber\\
 & \:\:\:\:\:\:\:\:\:\:\:\:\:\:\:
 \left.
 - j^{\tc{a}}_{(a)\!}(\mf{x})\phi^{(a)\!}(\mf{x})
 - j^{\tc{b}}_{(a)\!}(\mf{x})\phi^{(a)\!}(\mf{x})\right),
\end{align}
where the terms $\L_{\tc{a}}, \L_{\tc{b}},$ and $\L_{\phi}$ are the free Lagrangian densities of A,B and of the field $\phi$, respectively.

Stationarity of the action $S$ under variations of the field $\phi$ yields the equations of motion 
\begin{equation}\label{eq:EoMGeneral}
    \mathcal{P}[\phi_{(a)}] = j^{\tc{a}}_{(a)\!}+ j^{\tc{b}}_{(a)\!}, 
\end{equation}
where $\mathcal{P}$ is a differential operator determined by $\mathcal{L}_\phi$. 
We further assume that $\mathcal{P}$ is a linear differential operator, so that the solutions of Eq.~\eqref{eq:EoMGeneral} can be written in terms of retarded and advanced Green's functions.
The retarded Green's function $G_{R}^{(ab)\!}(\mf{x}, \mf{x}')$ is non-zero only when $\mf x$ is in the causal future of $\mf x'$ and the advanced $G_{A}^{(ab)\!}(\mf{x}, \mf{x}')$ when $\mf x$ is in the causal past of $\mf x'$. 
The two are related by $G_{A}^{(ab)\!}(\mf{x}, \mf{x}')=G_{R}^{(ba)\!}(\mf{x}', \mf{x})$. 
They generate solutions to the non-homogeneous equation $\mathcal{P}[\phi^{j}_{(a)}] = j_{(a)}$ as:
\begin{equation}\label{eq:FieldSourcedClassical}
    \phi^{(a)}_j(\mf x) = \int \dd V' G_{R/A}^{(ab)}(\mf x, \mf x')j_{(b)}(\mf x').
\end{equation}
Notice that a solution to $\mathcal{P}[\phi^{j}_{(a)}] = j_{(a)}$ is obtained when one considers any convex combination of the retarded and advanced Green's functions, however, these solutions always differ by a solution of the homogeneous equation $\mathcal{P}[\phi^{j}_{(a)}] = 0$. 
So, the field can always be reformulated in terms of the retarded propagator. 
Consequently, each system $\tc{I}\in\{\tc{A},\tc{B}\}$ sources a contribution to the scalar field:
\begin{equation}\label{eq:ClassicalFieldSourced}
    \phi^{(a)\!}_{\tc{i}} =  \int \!\dd V G_{R}^{(ab)\!}(\mf{x}, \mf{x}') j^{\tc{i}}_{(b)\!}( \mf{x}').
\end{equation}

Under the assumption that the field is entirely sourced by $\tc{A}$ and $\tc{B}$, one can compute the Hamiltonian of the system. We pick inertial coordinates $\mf{x}= (t,\bm{x})$, for which the volume element is $\dd V = \dd^{n+1} \mf{x}$ (where $n$ is the number of spatial dimensions), and replace the field by the solution $\phi^{(a)}(\mf x) = \phi^{(a)\!}_{\tc{a}}(\mf x)+\phi^{(a)\!}_{\tc{b}}(\mf x)$. The Hamiltonian generating evolution with respect to the inertial time parameter $t$ for the dynamics of the two systems takes the form
\begin{align}\label{eq:HamiltonianClassical}
 H(t) &= \!\int \!\dd^{n} \bm{x}  
 \bigg( \H_{\tc{a}} +\H_{\tc{b}} \nonumber\\
 & \:\:\:\:\:\:\:\:\:\:\:\:\:\:\:
 \left.
 +\frac{1}{2}j^{\tc{a}}_{(a)\!}(\mf{x})\phi^{(a)\!}(\mf{x})
 +\frac{1}{2}j^{\tc{b}}_{(a)\!}(\mf{x})\phi^{(a)\!}(\mf{x})\right),
\end{align}
where $\dd^{n} \bm{x}$ is the spatial volume element and $\H_{\tc{a}},\H_{\tc{b}}$ are the free Hamiltonian densities of A and B.
Notice that the interaction terms pick up a factor of $1/2$ coming from the free Hamiltonian density of the field. 
Physically, this can be understood as half of the potential energy sourced from A and B being stored in the field. A derivation of \eqref{eq:HamiltonianClassical} under an adiabatic approximation for the sources is provided in Appendix \ref{app:HamiltonianProof}, for the case of a scalar field. 
The interaction Hamiltonian for $\tc{A}$ and $\tc{B}$ can then be expressed entirely in terms of the sources:
\begin{align}\label{eq:HamiltonianInterClassical}
 H_{\rm{int}}(t) &= 
 \frac{1}{2}
 \!\int \!\dd V' \! 
 \!\int \!\dd^{3} \bm{x}\,
  G_{R}^{(ab)\!}(\mf{x}, \mf{x}') 
\nonumber\\
 & \:\:\:\:\:\:\:\:\:\:\:\:\:\:\:
 +\left(j^{\tc{a}}_{(a)\!}(\mf{x})j^{\tc{b}}_{(b)\!}(\mf{x}')
 +j^{\tc{a}}_{(b)\!}(\mf{x}')j^{\tc{b}}_{(a)\!}(\mf{x})\right),
\end{align}
where we neglected the self-interaction terms of the form $\frac{1}{2}j^{\tc{a}}_{(a)\!}(\mf{x})\phi^{(a)\!}_\tc{a}(\mf{x})$ and $\frac{1}{2}j^{\tc{b}}_{(a)\!}(\mf{x})\phi^{(a)\!}_\tc{b}(\mf{x})$. Eq.~\eqref{eq:HamiltonianInterClassical} is a direct coupling  between the classical sources, mediated by a relativistic potential. Overall, the formulation of the interaction between systems A and B in terms of the Hamiltonian of Eq.~\eqref{eq:HamiltonianInterClassical} is valid under the following assumptions 1) the field is exclusively sourced via retarded propagation by the two interacting systems and no external sources affect the field, 2) there is no self-interaction for the sources and 3) the sources are slowly varying in time, so that an adiabatic approximation is valid. 

\subsubsection{Quantum sources interacting via classical propagation}\label{subsubsec:QCAnalysis}

The model above can be extended to the case where systems $\tc{A}$ and $\tc{B}$ are quantum. This gives rise to the quantum-controlled model (\textit{qc-model}) studied in~\cite{quantClass,ourBMV}.
The goal of the qc-model is to prescribe an interaction between two quantum sources that preserves the relativistic nature of the classical model. 
However, the qc-model  does not incorporate local quantum degrees of freedom for the mediating field. Rather the field is fully determined by the quantum sources, thus the name quantum-controlled field. 

Let us consider two quantum sources, A and B, described in Hilbert spaces $\Hil_{\tc{a}}$ and $\Hil_{\tc{b}}$. The internal dynamics of these systems are implemented by free Hamiltonians (generating translations with respect to $t$) $\hat{H}_\tc{a}$ and $\hat{H}_\tc{b}$, respectively. The systems also couple to a scalar field according to the description of Eq.~\eqref{eq:LagrangianClassical} with self-adjoint operator valued currents\footnote{ The current operators as they appear in the expressions are operators in $\Hil_{\tc{a}}\otimes\Hil_{\tc{b}}$ of the form $\hat{j}^{\tc{a}}_{(a)\!}(\mf x) = \hat{J}{}^{\tc{a}}_{(a)\!}(\mf x)\otimes \openone_{\tc{b}}$, $\hat{j}^{\tc{b}}_{(a)\!}(\mf x) = \openone_{\tc{a}}\otimes\hat{J}{}^{\tc{b}}_{(a)\!}(\mf x)$.} $\hat{j}^{\tc{a}}_{(a)\!}(\mf x)$, $\hat{j}^{\tc{b}}_{(a)\!}(\mf x)$, prescribed in the interaction picture. 
In this model, the mediating field is a spacetime dependent operator acting on the Hilbert space $\Hil_{\tc{a}}\otimes\Hil_{\tc{b}}$. 
Motivated by the classical model, the field will be sourced analogously to Eq.~\eqref{eq:ClassicalFieldSourced}:
\begin{align}\label{eq:QCFieldSourced}
    \hat{\phi}^{(a)\!}_{j, \tc{c}}(\mf x) =  \!\int \!\dd V' G_{R}^{(ab)\!}(\mf{x}, \mf{x}') \hat{j}_{(b)\!}( \mf{x}'),
\end{align}
where label $\tc{C}$ stands for `controlled' and $G_{R}^{(ab)\!}(\mf{x}, \mf{x}')$ is the retarded propagator of the differential operator $\mathcal{P}$ in \eqref{eq:EoMGeneral}. 
Notice that the ``quantum'' field operator $\hat{\phi}^{(a)}_{\tc{c}}$ acts on the Hilbert space of the sources, and it is devoid of any local degrees of freedom of its own. 

The interaction Hamiltonian for the qc-model can be expressed in terms of the sources in a way analogous to \eqref{eq:HamiltonianInterClassical}:
\begin{align}\label{eq:QCInteractionHamiltonian}
 \hat{H}_{\tc{c}}(t) =& \frac{1}{2}\!\int \!\dd^{3} \bm{x} \! \!\int \!\dd V' \,
 G_{R}^{(ab)\!}(\mf{x}, \mf{x}')\nonumber
 \\
 &
 \:\:\:\:\:\:\:\:\:\:\:\:\:\:\:
 \left(
 \hat{j}^{\tc{a}}_{(a)\!}(\mf{x}) \hat{j}^{\tc{b}}_{(b)\!}( \mf{x}')+\hat{j}^{\tc{b}}_{(a)\!}(\mf{x})\hat{j}^{\tc{a}}_{(b)\!}( \mf{x}')\right).
\end{align}
The evolution of sources in this model is, therefore, unitary, unlike what happens when the sources are coupled to a quantum field that possesses its own degrees of freedom, which can exchange quantum information with the sources rendering the time evolution of the partial state of the sources non-unitary.

In order to later compare the dynamics implemented by the Hamiltonian of Eq.~\eqref{eq:QCInteractionHamiltonian} to the case of two sources coupled through a fully featured quantum field, we consider an initial state of the joint system $\hat{\rho}_{0,\tc{ab}}$ of the two sources (a density operator in $\Hil_{\tc{a}}\otimes\Hil_{\tc{b}}$) and compute its time evolution under the assumption that the sources are proportional to a sufficiently small constant $\lambda$: $\hat j_\tc{a}(\mf x),\hat j_\tc{b}(\mf x)\propto \lambda$.
The system evolves with respect to the unitary operator:
\begin{align}\label{eq:QCUnitary1}
     \hat{U}_\tc{c} &= \mathcal{T}\exp\left(-\ii \int\!\! \dd t \, \hat{H}_{\tc{c}}(t)\right).
\end{align}
 For an initial state of the joint system $\hat{\rho}_{0,\tc{ab}}=\hat{\rho}_{0}$, the final state of the system can be written as
\begin{equation}\label{eq:UnitaryEvolution}
    \hat{\rho}_\tc{c} = \hat{U}_\tc{c} \hat{\rho}_{0}\hat U_\tc{c}^{\dagger}.
\end{equation}
The state $\hat{\rho}_{\tc{c}}$ can be computed perturbatively using the Dyson expansion of the unitary \eqref{eq:QCUnitary1}:
\begin{equation}\label{eq:QCUnitaryExpand}
    \hat{U}_\tc{c}= \hat{U}^{(0)}_\tc{c} + \hat{U}^{(2)}_\tc{c} + \hat{U}^{(4)}_\tc{c}+ \mathcal{O}(\lambda^6),
\end{equation}
    where
    \begin{align}
        \hat{U}_\tc{c}^{(0)} &= \openone,\nonumber\\
        \hat{U}_\tc{c}^{(2)} &=  - \ii \int_{-\infty}^{+\infty} \!\!\!\!\!\dd t\, \hat{H}_{\tc{c}}(t), \label{eq:QCUnitaryDyson}\\
        \hat{U}_\tc{c}^{(4)} &=(-\ii)^{2} \int _{-\infty}^{+\infty} 
          \!\!\! \dd t  \int_{-\infty}^{+\infty} 
          \!\!\!\dd t'  \, \Theta(t-t')
         \hat{H}_{\tc{c}}(t)
         \hat{H}_{\tc{c}}(t'),\nonumber
    \end{align}
    where $\Theta(u)$ denotes the Heaviside step function. The final state of the joint system is:
    \begin{align}\label{eq:QCDensityMatrixPerturbativeDyson}
        \hat\rho_\tc{c}= \hat\rho_0+\hat\rho^{(2)}_\tc{c}+\hat\rho^{(4)}_\tc{c}+\mathcal{O}(\lambda^6),
    \end{align}
    where
    \begin{align}
        \hat\rho_\tc{c}^{(2)} &= \hat{U}_\tc{c}^{(2)} \hat{\rho}_{0} +\hat{\rho}_{0}\hat{U}_\tc{c}^{(2)}{}^{\dagger},\label{eq:QCSecondOrderCorrection}\\
        \hat\rho_\tc{c}^{(4)} &= \hat{U}_\tc{c}^{(2)} \hat{\rho}_{0} \hat{U}_\tc{c}^{(2)}{}^{\dagger} +\hat{U}_\tc{c}^{(4)} \hat{\rho}_{0} +\hat{\rho}_{0}\hat{U}_\tc{c}^{(4)}{}^{\dagger}.
    \end{align}
    Notice that all terms are of even order in the expansion parameter $\lambda$, since the interaction Hamiltonian \eqref{eq:QCInteractionHamiltonian} is quadratic in the currents.
            
    The integral of the interaction Hamiltonian can be expressed through the symmetric propagator:
    \begin{align}\label{eq:QCInteractionHamiltonianIntegral}
     \int \! \dd t \,\hat{H}_{\tc{c}}(t) =& 
     \frac{1}{2}\!\int \!\dd V' \! \!\int \!\dd V \,
     \Delta^{(ab)\!}(\mf{x}, \mf{x}')\hat{j}^{\tc{a}}_{(a)\!}(\mf{x}) \hat{j}^{\tc{b}}_{(b)\!}( \mf{x}'),
    \end{align}
    where the symmetric\footnote{Symmetry of $\Delta^{(ab)\!}(\mf{x}, \mf{x}')$ follows from $G_{A}^{(ab)\!}(\mf{x}, \mf{x}') = G_{R}^{(ba)\!}(\mf{x}', \mf{x})$.} propagator is 
    \begin{equation}\label{eq:SymmetricPropagator}
        \Delta^{(ab)\!}(\mf{x}, \mf{x}')
        = G_{R}^{(ab)\!}(\mf{x}, \mf{x}') + G_{A}^{(ab)\!}(\mf{x}, \mf{x}').
    \end{equation}
    Furthermore, one can show that at least up to fourth order the evolution depends only on integrals of the symmetric propagator of the field. 

    There are cases where the unitary $\hat{U}_\tc{c}$ can be found non-perturbatively (see, e.g.,~\cite{Landulfo:2016,analytical}). For instance, in the specific case of time-independent sources (that is, for sources such that $[\hat{j}^{\tc{i}}_{(a)\!}(\mf{x}), \hat{H}_\tc{i}] =0$ for $\tc{I}\in \{\tc{A},\tc{B}\}$), the unitary \eqref{eq:QCUnitary1} can be found explicitly. In these cases, the commutator of the interaction Hamiltonian densities vanishes, i.e. $[\hat{\H}_{\tc{c}}(\mf x) ,\hat{\H}_{\tc{c}}(\mf x') ]=0$, and the unitary becomes
    \begin{align}\label{eq:QCUnitaryCommutingSources}
         \hat{U}_{\tc{c}} &=\exp\left(-\ii \int\!\! \dd t \, \hat{\H}_{\tc{c}}(t)\right)
         \\
         &= 
         \exp\left(- \frac{\ii}{2}\!\int \!\dd V' \! \!\int \!\dd V \,
        \Delta^{(ab)\!}(\mf{x}, \mf{x}')\hat{j}^{\tc{a}}_{(a)\!}(\mf{x}) \hat{j}^{\tc{b}}_{(b)\!}( \mf{x}')\right). \nonumber
    \end{align}



\subsection{QC from QFT}



At first sight, the qc-model is simply ``putting hats'' on the classical sources in a relativistic description of the interaction between two currents. 
However, as naive as it may seem, the model is also a good approximation of a fully quantum theory in many regimes. 
In this subsection, we analyze when the qc-model can be understood as a limit of a fully quantum field theoretic description of the interaction of the quantum systems A and B.

To understand the relationship of QFT with a description via a qc-model, it will be helpful to describe the propagation of information in quantum fields on a similar footing as it is manifest in the qc-model, that is, in terms of propagators.


\subsubsection{Quantum field propagators}\label{sec:QCfromQFT}



In quantum field theory propagation is encoded in $n$-point functions. 
In the context of QFT, it will be convenient to define the retarded and advanced propagators as bi-distributions with kernels given by $G_{R}^{(ab)\!}(\mf x, \mf x')$, $ G_{A}^{(ab)\!}(\mf x, \mf x')$.
Given a set $\mathcal{F}(\mathcal{M})$ of smooth and well behaved\footnote{One should consider tensor fields that vanish fast enough at infinity. For instance, one can consider smooth compactly supported fields, or more general sets of functions (e.g. in the scalar case, one could consider test functions defined in Schwartz space).} test tensor fields $f_{(a)}$, we define the bi-distributions $\mathrm{G}_{R/A}:\mathcal{F}(\mathcal{M})\times\mathcal{F}(\mathcal{M}) \to \mathbb{C}$ such that
\begin{align}\label{eq:RetAdvBidistrDef}
    \mathrm{G}_{R/A}(f,g)=  \int \! \dd V  \!  \int \! \dd V' 
    G_{R/A}^{(ab)\!}(\mf{x},\mf{x}')f_{(a)\!}(\mf{x})g_{(b)\!}(\mf{x}').
\end{align}
The sum and difference of the retarded and advanced Green's functions, $\mathrm{G}_{R}$ and $\mathrm{G}_{A}$, play an important role in quantum field theory. 
We define them as the bi-distributions $\rm E,\mf \Delta:\mathcal{F}(\mathcal{M})\times\mathcal{F}(\mathcal{M}) \to \mathbb{C}$:
\begin{align}
    \mathsf{\Delta}(f,g)&\coloneqq  \mathrm{G}_R(f,g) + \mathrm{G}_A(f,g)\label{eq:SymmetricBidistr},\\
    \mathrm{E}(f,g)&\coloneqq  \mathrm{G}_R(f,g) - \mathrm{G}_A(f,g)\label{eq:AntisymmetricBidistr},
\end{align}
with respective integral kernels $\Delta^{(ab)}(\mf x, \mf x')$ (the symmetric propagator) in Eq.~\eqref{eq:SymmetricPropagator}, and \mbox{$E^{(ab)}(\mf x, \mf x')= G_{R}^{(ab)}(\mf x, \mf x')-G_{A}^{(ab)}(\mf x, \mf x')$} (the causal propagator). 
The latter, $\mathrm{E}$, will allow us to formulate covariant canonical commutation relations in quantum field theory.

Consider a real bosonic quantum field $\hat{\phi}^{(a)\!}$, where $(a)$ denotes an arbitrary set of Lorentz indices with equation of motion given by Eq.~\eqref{eq:EoMGeneral}. The observables of the theory form a unital $*$-algebra $\mathcal{A}$ called the \textit{algebra of observables}. 
In the most common construction of the algebra $\mathcal{A}$ the field  is an operator-valued distribution $\hat{\Phi}:\mathcal{F}(\mathcal{M})\to \mathcal{A}$, where $\mathcal{F}(\mathcal{M})$ is a space of smooth and well behaved test tensor fields $f_{(a)}$. 
The observables of the theory are generated by the smeared field operators $\hat{\Phi}(f)$, which can be formally written as
\begin{equation}
    \hat{\Phi}(f) = \int\! \dd V \hat{\phi}^{(a)\!}(\mf{x}) f_{(a)\!}(\mf{x}).
\end{equation}
The smeared field operators satisfy the covariant canonical commutation relations
\begin{equation}\label{eq:CovariantCommRel}
    [\hat{\Phi}(f), \hat{\Phi}(g)] = -\ii\mathrm{E}(f,g)\openone,
\end{equation} 
where $ \mathrm{E}(f,g)$ is defined in~\eqref{eq:AntisymmetricBidistr}.

In order to define the two-point functions of field observables, it is necessary to take expectation values.
These are taken with respect to some state of the system.
A state is defined as a linear functional that maps observables of the algebra to complex values $\omega: \mathcal{A}\to \mathbb{C}$. 
The choice of state induces a representation of the algebra $\mathcal{A}$ in a Hilbert space $\mathscr{H}$ through the GNS construction~\cite{Haag,Kasia}, where operators in the algebra correspond to linear operators acting on the Hilbert space. 
In specific cases, such as that of von Newman algebras of type I or Type II, the algebraic description of a state $\omega$ is related to a trace-class normalized positive self-adjoint density operator $\hat{\rho}_{\omega}\in \mathcal{B}(\mathscr{H})$, where $\mathcal{B}(\mathscr{H})$ is the set of bounded operators acting on $\mathscr{H}$, as follows:
\begin{equation}
    \text{For } \hat{ A} \in \mathcal{A}, \quad \omega(\hat{A}) = \langle \hat{A}\rangle_\omega= \tr(\hat{A}\hat{\rho}_\omega).
\end{equation}
We call $\omega(\hat A) = \langle \hat{A}\rangle_\omega$ the \textit{expectation value} of $\hat A$. 

 Two relevant two-point functions are the \textit{Wightman function} and the \textit{Feynman propagator}. 
The \textit{Wightman function} is defined as $\mathsf{W}:\mathcal{F}(\mathcal{M})\times\mathcal{F}(\mathcal{M})\to \mathbb{C}$ such that:
\begin{align}
    \mathsf{W}(f,g) &= \omega(\hat{\Phi}(f) \hat{\Phi}(g)) \nonumber\\
    &=\int \! \dd V  \!  \int \! \dd V'\, 
    W^{(ab)\!}(\mf{x},\mf{x}')f_{(a)\!}(\mf{x})g_{(b)\!}(\mf{x}'),
\end{align} 
with integral kernel $W^{(ab)\!}(\mf x, \mf x') = \langle \hat{\phi}^{(a)}(\mf{x}) \hat{\phi}^{(b)}(\mf x') \rangle_\omega$.
The time-ordered two-point function \mbox{$G_{F}^{(ab)\!}(\mf x, \mf x') = \langle \mathcal{T} \hat{\phi}^{(a)}(\mf{x}) \hat{\phi}^{(b)}(\mf x') \rangle_\omega$} defines the so-called \textit{Feynman propagator}, alternatively expressed as:
\begin{equation}\label{eq:FeynWightRelat}
    G_{F}^{(ab)\!}(\mf{x},\mf{x}') = \theta(t-t')W^{(ab)\!}(\mf{x},\mf{x}') + \theta(t'-t)W^{(ba)\!}(\mf{x}',\mf{x}).
\end{equation}
This gives rise to the bi-distribution \mbox{$\mathsf{G}_{F}:\mathcal{F}(\mathcal{M})\times\mathcal{F}(\mathcal{M})\to \mathbb{C}$}:
\begin{align}
    \mathsf{G_{F}}(f,g) = \int \! \dd V  \!  \int \! \dd V' \,
    G_{F}^{(ab)\!}(\mf{x},\mf{x}')f_{(a)\!}(\mf{x})g_{(b)\!}(\mf{x}').
\end{align}

We can see how the different propagators are related by splitting the kernels of the Wightman function and the Feynman propagator into their real and imaginary parts
\footnote{Using that $W^{(ab)\!}(\mf x, \mf x')^* = W^{(ba)}(\mf x', \mf x)$, one can obtain the real and imaginary parts of the two-point function from the decomposition $$\hat{\phi}^{(a)\!}(\mf{x})\hat{\phi}^{(b)\!}(\mf{x}') = \tfrac{1}{2}\{\hat{\phi}^{(a)\!}(\mf{x}),\hat{\phi}^{(b)\!}(\mf{x}')\} + \tfrac{1}{2}[\hat{\phi}^{(a)\!}(\mf{x}),\hat{\phi}^{(b)\!}(\mf{x}')].$$}.
\begin{align}
    W^{(ab)\!}(\mf{x}, \mf{x}') 
    &=\frac{1}{2}H^{(ab)\!}(\mf{x}, \mf{x}') - \frac{\ii}{2}E^{(ab)\!}(\mf{x}, \mf{x}'),\label{eq:WightmanSplit} \\
    G^{(ab)\!}_{F}(\mf{x}, \mf{x}') 
    &= \frac{1}{2}H^{(ab)\!}(\mf{x}, \mf{x}') + \frac{\ii}{2}\Delta^{(ab)\!}(\mf{x}, \mf{x}').\label{eq:FeynmanSplit}
\end{align}
Both propagators share the same real part, as can be shown from taking the real part of \eqref{eq:FeynWightRelat}. 
\mbox{$H^{(ab)\!}(\mf{x}, \mf{x}') = \langle \{\hat{\phi}^{(a)\!}(\mf{x}),\hat{\phi}^{(b)\!}(\mf{x}')\}\rangle_{\omega}$} defines the so-called \textit{Hadamard distribution}. 
It contains all the state dependence of both the Wightman function and the Feynman propagator, as both imaginary parts, $\Delta^{(ab)\!}(\mf{x}, \mf{x}')$, $ E^{(ab)\!}(\mf{x}, \mf{x}')$ are state independent due their definition through the retarded and advanced propagators. 
The imaginary parts have support when $\mf{x}$ is either in the causal future or the causal past of $\mf x'$. 

We note that the only propagator that takes part in the qc-model is the imaginary part of the Feynman propagator, as can be seen from Eq.~\eqref{eq:QCInteractionHamiltonianIntegral}). In what follows, we will see how these propagators contribute to the signalling between two quantum systems that locally couple to both a quantum field and a qc-field. 

\subsubsection{Local interactions with a quantum field} \label{subsubsec:QFTInter}
Now let us discuss the role of the propagators when two quantum systems A and B locally interact with a quantum field $\hat{\phi}^{(a)}$.  In analogy to the classical interaction~\eqref{eq:HamiltonianClassical}, we prescribe the interaction Hamiltonian as
\begin{align}\label{eq:InteractionHamQFT}
 \hat{H}_{\rm{int}}(t) = \lambda \!\int \!\dd^{3} \bm{x} 
 \left(
 \hat{j}^{\tc{a}}_{(a)\!}(\mf{x})\hat{\phi}^{(a)\!}(\mf{x})
 +\hat{j}^{\tc{b}}_{(a)\!}(\mf{x})\hat{\phi}^{(a)\!}(\mf{x})\right).
\end{align}
We will refer to this model as \textit{QFT model} throughout this paper.

Let us compute the dynamics of two quantum systems that evolve according to the Hamiltonian~\eqref{eq:InteractionHamQFT}.
We perform perturbative computations, similar to Subsection~\ref{subsubsec:QCAnalysis}. 
The main difference is that now the field has degrees of freedom described in a Hilbert space $\Hil_{\phi}$, thus the full quantum system is tripartite, consisting of subsystems $\tc{A}$, $\tc{B}$ and the field. The full system is then described in the Hilbert space $\Hil_{\tc{a}}\otimes\Hil_{\tc{b}}\otimes\Hil_{\phi}$.
We work in the interaction picture, where the state joint system evolves with respect to the time-evolution operator
\begin{align}\label{eq:QFTUnitary}
     \hat{U} &= \mathcal{T}\exp\left(-\ii \int\!\! \dd t \, \hat{H}_\text{int}(t)\right). 
\end{align}
Given an initial state\footnote{We will assume that we are working in a case where density operators for the field can be defined.} $\hat{\rho}_{0} \in \Hil_{\tc{a}}\otimes\Hil_{\tc{b}}\otimes\Hil_{\phi}$, the evolved state is given by
\begin{equation}\label{eq:UnitaryEvolutionQFT}
    \hat{\rho} = \hat U \hat{\rho}_{0} \hat U^{\dagger}
\end{equation}
During time evolution the two systems A and B may get entangled between themselves and with the field. The final state of the joint system AB is given by the reduced density matrix $\hat{\rho}_{\tc{ab}} = \Tr_{\phi}\hat{\rho}$.
From the Dyson expansion we get up to second order for the evolution operator:
    \begin{align}
     &\hat{U}= \hat{U}^{(0)} + \hat{U}^{(1)} + \hat{U}^{(2)}+ \mathcal{O}(\lambda^3),\label{eq:QFTUnitaryExpand} \\
        &\hat{U}^{(0)} = \openone\nonumber \\
        &\hat{U}^{(1)} =  - \ii \int_{-\infty}^{+\infty} \!\!\!\dd t\, \hat{H}_\text{int}(t)\nonumber\\
        &\hat{U}^{(2)} 
          =(-\ii)^{2} \int _{-\infty}^{+\infty} 
          \!\!\! \dd t  \int_{-\infty}^{+\infty} 
          \!\!\!\dd s  \, \Theta(t-t')
         \hat{H}_\text{int}(t)
         \hat{H}_\text{int}(s)\nonumber
    \end{align}
    and for the density matrix:
    \begin{align}
        &\hat\rho = \hat\rho_0+\hat\rho^{(1)}+\hat\rho^{(2)}+\mathcal{O}(\lambda^6)\\
        &\hat{\rho}^{(1)} = \hat{U}^{(1)}\hat{\rho}_{0}+\hat{\rho}_{0} +\hat{U}^{(1)}{}^{\dagger}\\
        &\hat\rho^{(2)} = \hat{U}^{(1)}\hat{\rho}_{0}\hat{U}^{(1)}{}^{\dagger}+\hat{U}^{(2)}\hat{\rho}_{0}+\hat{\rho}_{0}\hat{U}^{(2)}{}^{\dagger}.
    \end{align}

    \subsubsection{Comparison with the quantum-controlled model}
    
    Up to this point we have presented the interaction between two quantum systems A and B through quantum-controlled fields in \ref{subsec:QuantumControlled} and quantum fields in \ref{subsubsec:QFTInter}. 
    Now we compare the evolution of the state of the system AB up to second order in $\lambda$ and discuss the requisites for qc-fields to be a good approximation of quantum fields within this setup.
    
    Let us start with the QFT model, where the state of the stystem $\tc{AB}$ is the reduced state $\hat{\rho}_{\tc{ab}} =\Tr_{\phi}(\hat{\rho})$. 
    To make a clean comparison with the qc-model, we assume that the initial state of the field is uncorrelated with the state of the systems AB, i.e. the initial state is of the form $\hat{\rho}_{0} = \hat{\rho}_{\tc{ab},0}\otimes \hat{\rho}_{\phi}$.
    For simplicity, assume that the initial state of the field, $\hat{\rho}_{{\phi}}$, is quasi-free (zero-mean Gaussian), so that its odd-point functions vanish.

    The first order corrections of the reduced state $\hat{\rho}_{\tc{ab}} =\Tr_{\phi}(\hat{\rho})$, will be proportional to the expectation value of a single field operator, $\Tr_{\phi}(\hat{\phi}(\mf{x})\hat{\rho}_\phi)$, which vanishes due to the quasi-free assumption.

    The second order terms  $\hat{\rho}_{\tc{ab}}^{(2)} = \Tr_{\phi}(\hat{\rho}^{(2)})$ are hence of leading order and depend linearly on the two-point functions $ W^{(ab)\!}(\mf x, \mf x') = \Tr\left(\hat{\phi}^{(a)\!}(\mf{x})\hat{\phi}^{(b)\!}(\mf{x}')\hat{\rho}_{\phi}\right)$.
    To facilitate the comparison between the two models, we split $W^{(ab)\!}(\mf x, \mf x')$ into its real and imaginary parts, as in Eq.~\eqref{eq:WightmanSplit}. 
    Due to linearity, $\hat{\rho}_{\tc{ab}}^{(2)}$ also splits into two respective parts, namely $\hat{\rho}_{\tc{ab}}^{(2)} = \hat{\rho}_{\tc{ab}, \Re}^{(2)}+\hat{\rho}_{\tc{ab}, \Im}^{(2)}$. 
    
    The first term is expressed through the (kernel of the) Hadamard distribution as follows:
    \begin{align}
   \hat{\rho}_{\tc{ab}, \Re}^{(2)} &= \frac{\lambda^2}{2} \int \! \dd V  \!  \int \! \dd V'  H^{(ab)\!}(\mf x, \mf x')\label{eq:QFTRealCorrection}\\
        &\left(\hat{j}^{\tc{a}}_{(a)\!}(\mf{x}) \hat{\rho}_{\tc{ab},0} \hat{j}^{\tc{a}}_{(b)\!}(\mf{x}') 
        +\hat{j}^{\tc{b}}_{(a)\!}(\mf{x}) \hat{\rho}_{\tc{ab},0} \hat{j}^{\tc{b}}_{(b)\!}(\mf{x}')\right.\nonumber\\
        &+\hat{j}^{\tc{a}}_{(a)\!}(\mf{x}) \hat{\rho}_{\tc{ab},0} \hat{j}^{\tc{b}}_{(b)\!}(\mf{x}')
        +\hat{j}^{\tc{b}}_{(a)\!}(\mf{x}) \hat{\rho}_{\tc{ab},0} \hat{j}^{\tc{a}}_{(a)\!}(\mf{x}')\nonumber\\
        &-\hat{j}^{\tc{a}}_{(a)\!}(\mf{x}) \hat{j}^{\tc{b}}_{(b)\!}(\mf{x}') \hat{\rho}_{\tc{ab},0}
        -\hat{\rho}_{\tc{ab},0}\hat{j}^{\tc{a}}_{(a)\!}(\mf{x}) \hat{j}^{\tc{b}}_{(b)\!}(\mf{x}') \nonumber\\
        &\!-\!\Theta(t-t')\!\!\left(\hat{j}^{\tc{a}}_{(a)\!}(\mf{x}) \hat{j}^{\tc{a}}_{(b)\!}(\mf{x}') \hat{\rho}_{\tc{ab},0}
        +\hat{j}^{\tc{b}}_{(a)\!}(\mf{x}) \hat{j}^{\tc{b}}_{(b)\!}(\mf{x}') \hat{\rho}_{\tc{ab},0}\right) \nonumber\\
        &\!-\!\Theta(t'\!-t)\!\!\left.\left(\hat{\rho}_{\tc{ab},0}\hat{j}^{\tc{a}}_{(a)\!}(\mf{x}) \hat{j}^{\tc{a}}_{(b)\!}(\mf{x}')
        +\hat{\rho}_{\tc{ab},0}\hat{j}^{\tc{b}}_{(a)\!}(\mf{x}) \hat{j}^{\tc{b}}_{(b)\!}(\mf{x}')\right) \!\right)\!, \nonumber
    \end{align}
    where we used $\hat{j}^{\tc{a}}_{(a)}(\mf{x}) = \hat{j}^{\tc{a}\,\dagger}_{(a)}(\mf{x})$.
    The second term, that has the the imaginary part of the Wightman function, only depends on the retarded and advanced propagators, and can be organised as:
    \begin{align}
   &\hat{\rho}_{\tc{ab}, \Im}^{(2)} = \ii \frac{\lambda^2}{2} \int \! \dd V  \!  \int \! \dd V' \label{eq:QFTImCorrection}\\
        & -E^{(ab)\!}(\mf x, \mf x') 
        \left(\hat{j}^{\tc{a}}_{(a)\!}(\mf{x}) \hat{\rho}_{\tc{ab},0} \hat{j}^{\tc{a}}_{(b)\!}(\mf{x}') 
        +\hat{j}^{\tc{b}}_{(a)\!}(\mf{x}) \hat{\rho}_{\tc{ab},0} \hat{j}^{\tc{b}}_{(b)\!}(\mf{x}')\right.\nonumber\\
        &\:\:\:\:\:\:\:\:\:\:\:\:\:\:\:\:\:\:\:\:\:\:\:\:\left.+\hat{j}^{\tc{a}}_{(a)\!}(\mf{x}) \hat{\rho}_{\tc{ab},0} \hat{j}^{\tc{b}}_{(b)\!}(\mf{x}')
        +\hat{j}^{\tc{b}}_{(a)\!}(\mf{x}) \hat{\rho}_{\tc{ab},0} \hat{j}^{\tc{a}}_{(b)\!}(\mf{x}')\right)\nonumber\\
        &-G_{R}^{(ab)\!}(\mf x, \mf x')\left(\hat{j}^{\tc{a}}_{(a)\!}(\mf{x}) \hat{j}^{\tc{a}}_{(b)\!}(\mf{x}') \hat{\rho}_{\tc{ab},0}
        +\hat{j}^{\tc{b}}_{(a)\!}(\mf{x}) \hat{j}^{\tc{b}}_{(b)\!}(\mf{x}')\hat{\rho}_{\tc{ab},0}\right)\nonumber \\
        &+G_{A}^{(ab)\!}(\mf x, \mf x')\left( \hat{\rho}_{\tc{ab},0} \hat{j}^{\tc{a}}_{(a)\!}(\mf{x}) \hat{j}^{\tc{a}}_{(b)\!}(\mf{x}')
        +\hat{\rho}_{\tc{ab},0}\hat{j}^{\tc{b}}_{(a)\!}(\mf{x}) \hat{j}^{\tc{b}}_{(b)\!}(\mf{x}')\right)\nonumber \\
        &-\Delta^{(ab)\!}(\mf x, \mf x')[\hat{j}^{\tc{a}}_{(a)\!}(\mf{x})\hat{j}^{\tc{b}}_{(a)\!}(\mf{x}'),\hat{\rho}_{\tc{ab},0}]\nonumber.
    \end{align}
    On the other hand, from the qc-model, the evolution of the state up to second order can be calculated from \eqref{eq:QCSecondOrderCorrection} and \eqref{eq:QCUnitaryDyson}, yielding 
    \begin{align}\label{eq:QCsecondCorrection}
        \hat{\rho}_{\tc{c}}^{(2)} &= 
     -\ii \frac{\lambda^2}{2}\!\int \!\dd V' \! \!\int \!\dd V \,\Delta^{(ab)\!}(\mf{x}, \mf{x}')[\hat{j}^{\tc{a}}_{(a)\!}(\mf{x})\hat{j}^{\tc{b}}_{(b)\!}(\mf{x}'),\hat{\rho}_{\tc{ab},0}].
    \end{align}

    It is evident from  Eq.~(\ref{eq:QFTRealCorrection}-\ref{eq:QCsecondCorrection}) that the leading order time evolution in the qc-model corresponds to the last term of \eqref{eq:QFTImCorrection}. 
    Consequently, when this term is dominant, the qc-model is an approximation for the QFT description at leading order in perturbation theory. 
    That is for the system to be well modelled by the qc interaction, the following terms must be negligible:
    \begin{enumerate}
        \item The Hadamard contribution in Eq.~\eqref{eq:QFTRealCorrection},
        \item The asymmetric self-interaction terms in~\eqref{eq:QFTImCorrection},
        \item The antisymmetric causal propagator terms in~\eqref{eq:QFTImCorrection}. \label{ass:3}
    \end{enumerate}
    
    Indeed, in \cite{quantClass} it was shown that there are regimes where the conditions above are fulfilled.
    Concretely, it was found that when the interaction between the systems lasts for sufficiently long times, the systems are in causal contact with each other, and the coupling with the field is weak enough, the two models give the same predictions.

    Notice that in spacelike separation the leading order correction of the qc-model vanishes completely and the only term that survives in the QFT case is the Hadamard term in Eq. \eqref{eq:QFTRealCorrection}.
    Hence, interactions when in spacelike separation are well out of the regimes where the qc-model approximates the full QFT model.
    Indeed, the term containing the Hadamard in the reduced state for system $\tc{A}$ in Eq.~\eqref{eq:QFTRealCorrection} consists of terms that depend on local observables of $\tc{A}$ and the local noise due to the coupling with the field (encoded in $H^{(ab)}(\mf x, \mf x')$), i.e. 
    \begin{align}
   \hat{\rho}_{\tc{a}, \Re}^{(2)} &= \frac{\lambda^2}{2} \int \! \dd V  \!  \int \! \dd V'\,  H^{(ab)\!}(\mf x, \mf x') \nonumber\\
        &\left(\hat{j}^{\tc{a}}_{(a)\!}(\mf{x}) \hat{\rho}_{\tc{a},0} \hat{j}^{\tc{a}}_{(b)\!}(\mf{x}') \right.\nonumber\\
        &\:\:\:\:-\Theta(t-t')\hat{j}^{\tc{a}}_{(a)\!}(\mf{x}) \hat{j}^{\tc{a}}_{(b)\!}(\mf{x}') \hat{\rho}_{\tc{a},0}\nonumber\\
        &\:\:\:\:\left.-\Theta(t'-t)\hat{\rho}_{\tc{a},0}\hat{j}^{\tc{a}}_{(a)\!}(\mf{x}) \hat{j}^{\tc{a}}_{(b)\!}(\mf{x}') \right).\label{eq:QFTReCorrection2}
    \end{align}
This means that the qc-model cannot capture the fact that spacelike separated systems coupled locally to the field can become correlated even in spacelike separation by extracting previously existing field correlations (see, for instance~\cite{Valentini1991,Reznik1,reznik2,Salton:2014jaa,Pozas-Kerstjens:2015,Pozas2016,HarvestingSuperposed,Henderson2019,bandlimitedHarv2020,ampEntBH2020,mutualInfoBH,threeHarvesting2022,twist2022,ericksonNew}). 

Of course the qc-model is just one possible model where relativistic aspects are captured while the field does not have any degrees of freedom. One could then wonder whether it is possible to make the qc-model more accurate by modifying it to incorporate more terms present in the QFT description (Eqs.~\eqref{eq:QFTRealCorrection} and \eqref{eq:QFTImCorrection}), still without introducing local degrees of freedom for the field. 
The Hadamard part in Eq. \eqref{eq:QFTRealCorrection} cannot be reproduced by a unitary evolution compatible with relativistic locality, as it describes noise localised around the quantum sources and spacelike correlations between them.
Attempting to incorporate the non-local Hadamard terms without a mediating field would interfere with relativistic locality, introducing an instantaneous interaction between the sources\footnote{In QFT correlations between spacelike separated systems can be acquired due correlations previously existing in the field and instantaneous interactions are avoided due to the microcausality condition, which demands that the field observables commute in spacelike separation.}.
The terms from Eq.\eqref{eq:QFTImCorrection} that are asymmetric or antisymmetric are also incompatible with the unitary evolution of states. 
Indeed, the leading order evolution of a state with respect to a unitary of the form of Eq.~\eqref{eq:QCUnitary1} can only yield terms that depend on the commutator $[H_{\tc{c}}(t),\hat{\rho}_0]$.
The anti-symmetric terms in Eq.\eqref{eq:QFTImCorrection} cannot be brought to this form, since the currents are present in pairs in the Hamiltonian \eqref{eq:QCInteractionHamiltonian}. 
This leads to the conclusion that the only possible change to the model could be an inclusion of asymmetric self-interaction terms that can reproduce the retarded and advanced terms in Eq. \eqref{eq:QFTImCorrection}, but not the Hadamard or antisymmetric terms. 
We leave such an study for future work.

\section{Causality in effective QFT descriptions}\label{sec:Retrocausal}

As we saw in the previous section, the qc-model can be thought of as a limit of a QFT description in specific regimes. 
Consequently, there is no reason to expect the qc-model to be physical outside its regimes of validity. 
For instance, the model might give predictions that are not fully compatible with relativistic principles, allowing for retrocausal signalling between the probes. 
In this section we will discuss possible causality violations that effective models can introduce in a general setup where two parties interact in distinct spacetime regions. We give an explicit definition of retrocausation for two systems that interact while they are localised in spacetime, and explain how its occurrence relates to the capacity of the qc-model to approximate QFT.

\subsection{Retrocausal signalling}\label{sub:RetrocausalSignalDef}

In this subsection, we present the conditions for a relativistic  model of field-probe interaction to display retrocausality in communication settings. 

Let us consider two systems A and B signalling each other via propagation through a field. 
The first ingredient of our analysis is defining the notion of retrocausal signalling. Consider two emitters A and B. 
Let A be in the causal past of B. In this case, system A can emit a signal to B but A cannot receive any signal back from B. 
Relativistic causality demands that A cannot have information on whether or not the signal eventually reached B. 
We will call such an effect \textit{retrocausal}.
We can then quantify retrocausality in terms of the strength of that effect, i.e. how much does the state of A depend on any information about B.
In essence, retrocausality implies the ability to obtain information about systems that lie outside of one's causal past.


To define retrocausal effects on quantum systems in spacetime, we will examine the dependence of the statistics of an emitter on the retarded propagation of signals.
We start by noticing that the spacetime localization of systems A and B can be encoded in sets of spacetime smearing functions $\{\Lambda_\tc{a}^i(\mf x)\}$ and $\{\Lambda_\tc{b}^i(\mf x)\}$~\footnote{For instance, in the qc-model and the QFT model defined by the respective Hamiltonians~\eqref{eq:QCInteractionHamiltonian} and~\eqref{eq:InteractionHamQFT}, the set of spacetime smearing functions that define system A would be given by $\Lambda^{nm}_\tc{a}(\mf x) = \bra{n_\tc{a}}\hat{j}_\tc{a}(\mf x) \ket{m_\tc{a}}$, where $\ket{n_\tc{a}}$ forms a basis for the Hilbert space of $\Hil_\tc{a}$.}. 
For a general choice of spacetime smearing functions, the state of A will depend on both retarded propagated signals from B to A (encoded in terms of the form $\mathrm{G}_{\tc{ab}}^{ij} = \mathrm{G}_{R}(\Lambda_{\tc{a}}^i,\Lambda_{\tc{b}}^j)$), and signals from A to B (encoded in $\mathrm{G}_{\tc{ba}}^{ji} = \mathrm{G}_{R}(\Lambda_{\tc{b}}^j,\Lambda_{\tc{a}}^i)$).
For simplicity, we abbreviate the dependencies on $\mathrm{G}_{\tc{ab}}^{ij}$ and $\mathrm{G}_{\tc{ba}}^{ji}$ by denoting them simply as $\mathrm{G}_{\tc{ab}}$ and $\mathrm{G}_{\tc{ba}}$. 
Using this notation, the dependence of the final state of A on the signals from A to B and B to A can be expressed as ${\hat{\rho}_{\tc{a}} = \hat{\rho}_{\tc{a}}(\mathrm{G}_{\tc{ab}},\mathrm{G}_{\tc{ba}})}$.

Within this notation, we will say that retrocausal effects appear when the state of A to depends on $\mathrm{G}_{\tc{ba}}$, when the relative causal ordering between A and B is such that B is in the causal future of A, i.e. for a relativistic field theory $\mathrm{G}_{\tc{ba}} = 0$.
Within this formalism we can now define the condition that a model must satisfy so that it does not allow retrocausal signalling between two compactly supported systems (that is $\Lambda_\tc{a}^i(\mf x)$ and $\Lambda_\tc{b}^i(\mf x)$ compactly supported in spacetime). \\

\begin{definition}\label{def:NonRetrCompact}
    Consider a model that defines the evolution between two compactly supported quantum systems $\tc{A}$ and $\tc{B}$ such that their respective statistics after the interaction is prescribed by the reduced states $\hat\rho_{\tc{a}}(\mathrm{G}_\tc{ab},\mathrm{G}_\tc{ba})$ and $\hat\rho_{\tc{b}}(\mathrm{G}_\tc{ab},\mathrm{G}_\tc{ba})$. We call the model \textbf{non-retrocausal} if
    \begin{equation}\label{eq:condnonRetro}
    \hat{\rho}_{\tc{a}}(0,\mathrm{G}_{\tc{ba}}) = \hat{\rho}_{\tc{a}}(0,0), \quad \forall \,\,\mathrm{G}_{\tc{ba}}.
    \end{equation}
    \textit{for any choices of $\Lambda_\tc{a}^i(\mf x)$ and $\Lambda_\tc{b}^i(\mf x)$ such that $G_\tc{ab} = 0$.}
\end{definition}

\begin{figure}[h]
        \centering
        \includegraphics[width=0.5\linewidth]{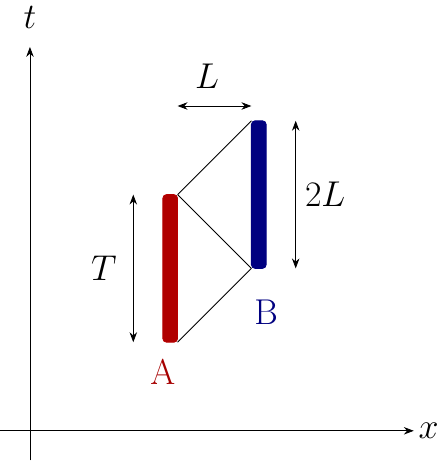}
        \caption{A spacetime diagram of the interaction of systems A (red) and B (blue) in two compactly supported regions, where A can signal to B, but B cannot signal to A.}
        \label{fig:OnlyRetrocausal}
    \end{figure}

Intuitively, the non-retrocausality condition above states that in a non-retrocausal model the reduced state of A cannot depend on signals that were received by B, if B does not signal to A.
One scenario where $\mathrm{G}_{\tc{ab}}= 0$ and $\mathrm{G}_{\tc{ba}} \neq 0$ is presented in Fig.~\ref{fig:OnlyRetrocausal}, where system A is supported only in the causal past of B. 
In this case, if $\hat{\rho}_{\tc{a}}$ depended on any information from B (encoded in the dependence on $\textrm{G}_{\tc{ba}}$), the model would be retrocausal. 


For instance, the QFT model does satisfy Def.~\ref{def:NonRetrCompact}, thus it is non-retrocausal. 
To demonstrate this to leading order, we can use the results of Subsection~\ref{subsubsec:QFTInter}, discussing the dependence of the state of A on system B after their interaction, when they are initially uncorrelated. 
These terms come from Eq.~\eqref{eq:QFTImCorrection} and the reduced state of A reads
\begin{align}
   \hat{\rho}_{\tc{a}, \Im}^{(2)} = - \ii \lambda^2 \int \! \dd V  \!  \int \! \dd V'  G_{R}^{(ab)\!}(\mf x, &\mf x') [\hat{j}^{\tc{a}}_{(a)\!}(\mf{x}),\hat{\rho}_{\tc{a},0}] \nonumber\\& \Tr{\hat{j}^{\tc{b}}_{(b)\!}(\mf{x}')\hat{\rho}_{\tc{b},0}}.\label{eq:QFTImCorrection48}
    \end{align}
Notice that the expression above depends exclusively on the retarded propagation from B to A, which vanishes whenever B is in the causal future of A (that is, whenever $\mathrm{G}_\tc{ab} = 0$). 
Consequently, the QFT model does not allow for retrocausal effects.

On the other hand, the qc-model can be retrocausal. 
This can be seen from Eq.~\eqref{eq:QCsecondCorrection} in a case like, for example, that of Fig.~\ref{fig:OnlyRetrocausal}, where $\mathrm{G}_{\tc{ab}}=0$. In this scenario, for the qc-model, the state of A depends non-trivially on $\mathrm{G}_\tc{ba}\neq 0$:
\begin{align}\label{eq:RetrocausalContrQC}
        \hat{\rho}_{\tc{a},\tc{c}}^{(2)} = 
     -\ii \frac{\lambda^2}{2}\!\int \!\dd V' \! \!\int \!\dd V \,G_{A}^{(ab)\!}(\mf{x}, &\mf{x}')[\hat{j}^{\tc{a}}_{(a)\!}(\mf{x}),\hat{\rho}_{\tc{a},0}]  \nonumber\\& \Tr{\hat{j}^{\tc{b}}_{(b)\!}(\mf{x}')\hat{\rho}_{\tc{b},0}},
\end{align}
thus, the qc-model does not satisfy Def.~\ref{def:NonRetrCompact}.

Now that we have a characterisation of whether a model allows for retrocausal signalling or not, we will define when a particular interaction scenario is retrocausal. 
Even if a model can allow for retrocausal signalling, this does not necessarily imply that every possible setup within the model will suffer from it. 
It may be possible to devise setups in retrocausal models where retrocausal effects are not present. 
A very simple example is the scenario in Fig.~\ref{fig:OnlyRetrocausal} when we look at the dependence of B on A. In this case there would be no retrocausal effect on B, since A is in its causal past and any change of B due to A would be compatible with relativistic causality. 
Below, we define when the state of A is retrocausally affected by B in a particular setup.



\begin{figure}[h]
        \centering
        \includegraphics[width=0.5\linewidth]{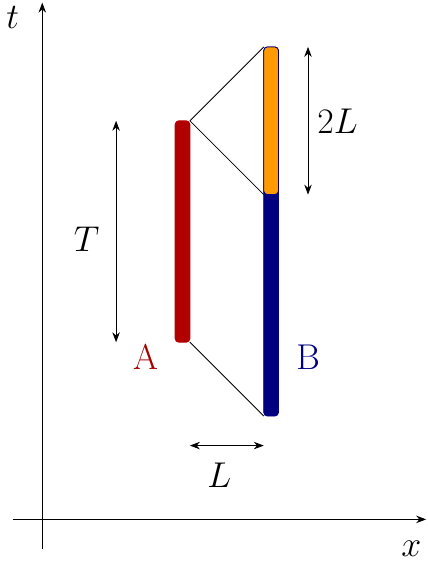}
       \caption{A spacetime diagram of the interaction regions of systems A and B separated by a distance $L$ when A has support in the future lightcone of B. The time duration of the interaction of systems A is $T$ and the time duration of B is $T+2L$. The region in blue signals causally to A and the orange region is associated to retrocausal signalling from B to A.}
		\label{fig:setup}
    \end{figure}

\begin{definition}\label{def:NonRetrCaseCompact}
    Consider two quantum systems, $\tc{A}$ and $\tc{B}$, supported in regions $\mathcal{R}_{\tc{a}}$ and $\mathcal{R}_{\tc{b}}$ in spacetime. Their localization is implemented by sets of compactly supported functions $\{\Lambda_\tc{a}^i(\mf x)\}$ and $\{\Lambda_\tc{b}^i(\mf x)\}$.
    Their respective states after the interaction can be written as $\hat\rho_{\tc{a}}(\mathrm{G}_\tc{ab},\mathrm{G}_\tc{ba})$ and $\hat\rho_{\tc{b}}(\mathrm{G}_\tc{ab},\mathrm{G}_\tc{ba})$.
    Let $\Tilde{\mathcal{R}}_{\tc{b}}$ denote a subregion of $\mathcal{R}_{\tc{b}}$ with $\Tilde\Lambda^i_\tc{b}(\mf x)=\left.\Lambda^i_\tc{b}(\mf x)\right|_{\tilde{\mathcal{R}}_\tc{b}}$ being the restriction of $\Lambda_\tc{b}^i(\mf x)$ to the region $\tilde{\mathcal{R}}_\tc{b}$ and  $\Tilde{\mathrm{G}}_{\tc{ab}}$ the retarded propagator evaluated at ${\Lambda}^i_\tc{a}(\mf x)$ and $\tilde{\Lambda}^i_\tc{b}(\mf x)$. 
    We then say that \textbf{the setup has no retrocausal effect on A} if either\\
    
    \noindent 1. There does not exist a subregion $\Tilde{\mathcal{R}}_{\tc{b}}\subset\mathcal{R}_{\tc{b}}$ such that
        $$\Tilde{\mathrm{G}}_\tc{ab} =0\:\:\text{and}\:\: \Tilde{\mathrm{G}}_\tc{ba}\neq 0,$$
    or\\
    
    \noindent 2. There exists a subregion $\Tilde{\mathcal{R}}_{\tc{b}}\subset\mathcal{R}_{\tc{b}}$ such that
    \begin{align}
        \Tilde{\mathrm{G}}_\tc{ab} &=0\:\:\text{and}\:\: \Tilde{\mathrm{G}}_\tc{ba}\neq 0,\nonumber\\        
        \text{\textit{but}} \quad\hat{\rho}_{\tc{a}}(\mathrm{G}_{\tc{ab}},\mathrm{G}_{\tc{ba}}) &= \hat{\rho}_{\tc{a}}(\mathrm{G}_{\tc{ab}},\mathrm{G}_{\tc{ba}}-\Tilde{\mathrm{G}}_\tc{ba}).~\footnote{Notice that $\mathrm{G}_{\tc{ba}}-\Tilde{\mathrm{G}}_\tc{ba}$ is the smeared retarded propagator between the regions $\mathcal{R}_{\tc{a}}$ and $\mathcal{R}_{\tc{b}}\setminus\Tilde{\mathcal{R}}_{\tc{b}}$. This is due to the linear dependence of $ \mathrm{G}_{\tc{ba}}$ on the smearing functions.}
        \end{align}
\end{definition}
According to this definition, B does not retrocausally signal to A if either 1) B is  completely in the causal past of A, or 2) if there is a subregion of B outside of the domain of dependence of A, but the final state of A does not depend on any information from B coming from this subregion. 
This is illustrated in Fig.~\ref{fig:setup}. 
In the orange region of duration $2L$, B can receive signals form A, but the part of B in the orange region should not be able to affect A in a non-retrocausal model. 
We can also reformulate this in terms of \text{reciprocation of signals}: In a non-retrocausal scenario A can signal to B in the orange region but B should not be able to \textit{reciprocate} and signal back from the orange region to Alice.

\subsection{Asymmetric signalling and retrocausality in the qc-model}

We will now explore the definitions of Subsection~\ref{sub:RetrocausalSignalDef} in the context of the qc-model and its ability to approximate an interaction described by fully relativistic QFT. 
In Subsection~\ref{subsubsec:QCAnalysis}, we saw that, up to leading order in perturbation theory, the final state of A and B in the qc-model depends only on terms containing the symmetric propagator $\Delta(\Lambda_\tc{a},\Lambda_\tc{b})$. That is, the final state of system A after the interaction has the form \mbox{$\hat{\rho}_\tc{a}(\textrm{G}_{\tc{ab}},\textrm{G}_{\tc{ba}}) = \hat{\rho}_\tc{a}(\textrm{G}_{\tc{ab}} + \textrm{G}_{\tc{ba}})$} (Eq.~\eqref{eq:SymmetricPropagator}). In general, using
\begin{equation}\label{eq:temp}
    \textrm{G}_{\tc{ab}} = \frac{1}{2}(\mf{\Delta}_{\tc{ab}} + \textrm{E}_{\tc{ab}}),\quad 
    \textrm{G}_{\tc{ba}} = \frac{1}{2}(\mf{\Delta}_{\tc{ab}} -\textrm{E}_{\tc{ab}}),
\end{equation}
the dependence of the final state can be rewritten in terms of the causal and symmetric propagators as $\hat{\rho}_\tc{a}(\textrm{G}_{\tc{ab}},\textrm{G}_{\tc{ba}}) = \hat{\rho}_\tc{a}(\textrm{E}_{\tc{ab}},\mf{\Delta}_{\tc{ab}})$. 
Then, it is more convenient to say that for time evolution described by the qc-model the state is independent of $\rm{E}_{\tc{ab}}$, i.e. $\hat{\rho}_\tc{a}(\textrm{E}_{\tc{ab}},\mf{\Delta}_{\tc{ab}}) = \hat{\rho}_\tc{a}(\mf{\Delta}_{\tc{ab}})$.


We will now show that in the qc-model, if $\rm{E}_{\tc{ab}}$ vanishes, it is not possible for system A to not receive retrocausal signals from B. 
For a non-retrocausal interaction setup, we can rewrite the condition of Def.~\ref{def:NonRetrCaseCompact} in terms of the causal and symmetric propagators:
\begin{equation}
        \hat{\rho}_{\tc{a}}(\mathsf{\Delta}_{\tc{ab}},\mathrm{E}_{\tc{ab}}) = \hat{\rho}_{\tc{a}}({\mathsf{\Delta}}_{\tc{ab}}- \Tilde{\mathrm{G}}_\tc{ba},{\mathrm{E}}_{\tc{ab}}+ \Tilde{\mathrm{G}}_\tc{ba}),\label{eq:Def2DeltaE}
    \end{equation}
where $\Tilde{\mathrm{G}}_\tc{ba}$ is the propagation from A to the region of B outside the domain of dependence of A. 
Assuming $ \tilde{\mathrm{G}}_\tc{ba}\neq 0,$ the terms $ {\mathrm{E}}_{\tc{ab}}+ \Tilde{\mathrm{G}}_\tc{ba}$ and $\mathrm{E}_{\tc{ab}}$ cannot vanish simultaneously. This in turn means that if $ \tilde{\mathrm{G}}_\tc{ba}\neq 0$, imposing $\mathrm{E}_{\tc{ab}}=0$ prevents Eq.~\eqref{eq:Def2DeltaE} from being satisfied, thus allowing for retrocausal signalling from B to A.  All this reasoning leads to an important conclusion:\\

\noindent\textit{The capacity of the qc-model to approximate QFT inescapably implies some degree (even if small) of retrocausal signalling.}\\


That is, there is a trade-off in the qc-model between retrocausality and the ability to approximate finite time interactions in QFT. This might seem paradoxical: QFT does not predict any causality violations and is well approximated by the qc-model exactly in the regimes where qc-interactions lead to retrocausal signals. This apparent paradox is resolved by noticing that although the qc-model always predicts retrocausal effects when it approximates QFT interactions, the strength of these retrocausal signals can, in principle, be arbitrarily small (e.g. when the interaction times are sufficiently long). 




We can see the tradeoff between approximating QFT and retrocausal effects in an explicit setup. For instance, consider the setup shown in Fig. \ref{fig:setup}. Without the orange region, system A cannot be retrocausally signalled, as B would only be supported in its causal past. 
In this case $\rm{G}_{\tc{ba}}\neq \rm{G}_{\tc{ab}}$, thus $\rm E_{\tc{ab}}\neq 0$.
The more we extend the temporal support of B to the future, the more $E_{\tc{ab}}$ decreases giving a better approximation to QFT. 
But in the qc-model, the orange region can signal to A retrocausally, showcasing the trade-off between retrocausality and approximating QFT.

Notice that, as mentioned above, the causality violations present in the qc-model do not imply that the qc-model is always a bad effective description for the interaction. This is because the strength of the  retrocausal effects depends on the temporal support of the two systems. In the next section we will quantify the causality violations introduced by the qc-model, showing that these become in fact arbitrarily small in the regimes where the qc-model approximates QFT.

\section{Quantifying retrocausality and regime of applicability of the qc-model}
\label{sec:quantify}

Now that we have established that the qc-model can be retrocausal, we would like to quantify the impact of retrocausal effects in the predictions of the model, and furthermore identify the regimes where the qc-model is still a good description of  physical interactions. We will analyze this quantitatively in the specific case of two-level systems interacting via a massless scalar qc-field, however notice that this analysis straightforwardly carries on to more general setups such as, for example, qc-models describing the gravitational interaction between two masses~\cite{ourBMV}.

To begin, let us consider two two-level quantum systems A and B coupled via a massless scalar qc-field. 
This setup is the qc-analogue to the so-called two-level Unruh-DeWitt detector model~\cite{Unruh-Wald,DeWitt} which has found many applications in the context of relativistic quantum information~\cite{Valentini1991,Reznik1,reznik2,Salton:2014jaa,teleportation2014,Jonsson2,collectCalling,Pozas-Kerstjens:2015,Pozas2016,teleportation,nichoTeleport,HarvestingSuperposed,Henderson2019,bandlimitedHarv2020,ampEntBH2020,Simidzija_2020,mutualInfoBH,threeHarvesting2022,twist2022,ericksonNew,KojiCapacity,KojiEntTeleport}.
The massless scalar field obeys the equations of motion: 
\begin{equation}\label{eq:EoMScalar}
        \Box \phi  = 0,
    \end{equation}
where, in inertial coordinates $(t,\bm x)$, $\Box = \partial_\mu\partial^\mu $.
We will be concerned with $3+1$ and $1+1$ dimensional Minkowski spacetimes. 
The quantum systems A and B are qubits that undergo inertial motion, commoving with the inertial time $t$ coordinate. 
The free dynamics of the qubits are implemented by the free Hamiltonians that generate time translations with respect to the inertial time $t$,
\begin{equation}\label{eq:FreeHamiltonian}
    \hat{H}_{\tc{i}} = \Omega_{\tc{i}} \hat{\sigma}^{+}_{\textsc{i}} \hat{\sigma}^{-}_{\textsc{i}},
\end{equation}
 where $\Omega_{\tc{i}}\geq0$ are their proper energy gaps and $\hat{\sigma}_{\textsc{i}}^{\pm}$ are $\mathfrak{su}(2)$ ladder operators.  
 
 Systems A and B couple to the scalar field through their monopole moments $\hat{m}_{\tc{i}}(t)$ associated with Hilbert spaces $\Hil_{\tc{i}}\cong \mathbb{C}^{2}$ for $\tc{I}\in \{\tc{A},\tc{B}\}$. 
In the total Hilbert space $\Hil_{\tc{ab}} = \Hil_\tc{a}\otimes\Hil_{\tc{b}}$, the monopole moments are expressed as $ \hat{\mu}_{\tc{a}}(t) = \hat{m}_{\tc{a}}(t)\otimes\openone_{\textsc{b}}$ and $\hat{\mu}_{\textsc{b}} (t)= \openone_{\textsc{a}}\otimes \hat{m}_{\textsc{b}}(t)$. 
The interactions of the two systems with the field are localised by the spacetime smearing functions $\Lambda_{\tc{a}}(\mf{x})$ and $\Lambda_{\tc{b}}(\mf{x})$. 
For this setup, the current densities that couple to the qc-field are:
\begin{equation}\label{eq:CurrentDensities}
    \hat{\!j}_{\tc{i}}(\mf{x}) = \Lambda_{\tc{i}}(\mf{x})\hat{\mu}_{\tc{i}}(t).    
\end{equation}
As usual in particle detector models, we assume the spacetime smearing functions to split into a product of spatial and temporal smearings in their own rest space, corresponding to the assumption of a Fermi-rigid system~\cite{us,generalPD}:
\begin{equation}\label{eq:SmearingChoice}
    \Lambda_{\tc{i}}(\mf{x}) =  
    \chi_{\tc{i}}(t)F_{\tc{i}}(\bm{x}), 
\end{equation}
where $F_{\tc{i}}(\bm{x})$ is the spatial smearing and $\chi_{\tc{i}}(t)$ is the switching function.

We will assume that systems A and B are localised by spacetime smearing functions $\Lambda_{\tc{a}}$ and $\Lambda_{\tc{b}}$ of compact support in spacetime regions $\mathcal{R}_{\tc{a}}$ and $\mathcal{R}_{\tc{b}}$.  
Then, it is possible to split $\mathcal{R}_{\tc{b}}$ into two parts, which either emit a signal to A or not, that we call $\mathcal{R}_{\tc{b}}^{(c)}$ and $\mathcal{R}_{\tc{b}}^{(r)}$. 
The parts of the support of B that emit signal to A will strongly depend on several factors such as whether the field is massless or not, the spacetime geometry, and the number of spatial dimensions\footnote{For example, for a massless scalar field in 3+1 Minkowski spacetime, the strong Huygens principle~\cite{Hyugens1,Huygens2,RayHyugens} applies and the field only propagates along null geodesics. However, even for a massless field in 1+1D the field can propagate along timelike hypersurfaces}. Considering all this, in a given setup we can split the spacetime smearing function $\Lambda_{\tc{b}}$ into two parts $\Lambda_{\tc{b}}^{(c)}=\left.\Lambda_{\tc{b}}\right|_{\mathcal{R}_{\tc{b}}^{(c)}} $ and $\Lambda_{\tc{b}}^{(r)}=\left.\Lambda_{\tc{b}}\right|_{\mathcal{R}_{\tc{b}}^{(r)} }$ and we can write: 
\begin{align}\label{eq:SplitSupport}
    \Lambda_{\tc{b}}  = \Lambda_{\tc{b}}^{(c)}+\Lambda_{\tc{b}} ^{(r)}.
\end{align}

In the following analysis, we will follow~\cite{EduCusality2015} and define a signalling estimator that quantifies the signalling from system B to system A. 
Using the split of Eq.~\eqref{eq:SplitSupport} we compute the signalling estimator from the regions $\mathcal{R}_{\tc{b}}^{(c)}$ and $\mathcal{R}_{\tc{b}}^{(r)}$ to A, allowing us to quantify the effects of retrocausal signalling. 
We start our analysis with the case of detectors with an energy gap, using perturbative results in (3+1) and (1+1) dimensions and later we present the case of gapless detectors, where the analysis is non-perturbative. 




\subsection{Quantum systems with non-trivial internal dynamics} \label{sub:Gap}

    Let us now initiate the analysis with the case of detectors with a gap, i.e. $\Omega_{\tc{i}}>0$, for which the detectors A and B have non-trivial internal dynamics. 
    We choose their initial states to be $\hat{m}_{\textsc{i}} (t=0) = \hat{\sigma}_{\textsc{i}}^{+}+ \hat{\sigma}_{\textsc{i}}^{-}$ such that they do not commute with the free Hamiltonian in Eq.~\eqref{eq:FreeHamiltonian}. 
    The time-evolved monopole moments of A and B in the interaction picture then read
    \begin{equation}
        \hat{m}_{\textsc{i}} (t) = \hat{\sigma}_{\textsc{i}}^{+} e^{\ii\Omega_{\tc{i}} t}+ \hat{\sigma}_{\textsc{i}}^{-}e^{-\ii\Omega_{\tc{i}} t}.
    \end{equation}
    The current densities $j_{\tc{i}}(\mf x)$ that define the coupling are given by Eq.~\eqref{eq:CurrentDensities}. 
    To leading order, the time evolution operator of the qc-interaction (Eq.~\eqref{eq:QCUnitaryDyson}) then reads
    \begin{align}
        \hat{U}_{\tc{c}}  & = \openone
        -
        \frac{ \ii \lambda^2}{2}\!\!
        \int\!\!
        \dd V   
        \dd V'   
        \Delta(\mf{x},\mf{x}') 
        \Lambda_{\tc{a}}(\mf{x}) 
        \Lambda_{\tc{b}}(\mf{x}') 
        \hat{\mu}_{\tc{a}}(t)\hat{\mu}_{\tc{b}}(t') ,
    \end{align}
    and the final state of the system (Eq.~\eqref{eq:QCSecondOrderCorrection}) is given by
    \begin{align}
        \hat\rho_{\tc{c}} 
        &=\hat{\rho}_{0}-\frac{\ii\lambda^2}{2}
        \!\!\int \!\!
        \dd V   
        \dd V'   
        \Delta(\mf{x},\mf{x}') 
        \Lambda_{\tc{a}}(\mf{x}) 
        \Lambda_{\tc{b}}(\mf{x}') \\
        & \:\:\:\:\:\:\:\:
         \:\:\: \:\:\: \:\:\:
       \times \Big[
        \hat{\mu}_{\tc{a}}(t)
        \hat{\mu}_{\tc{b}}(t')
        \hat{\rho}_{0} 
        -
        \hat{\rho}_{0}
        \hat{\mu}_{\tc{a}}(t')
         \hat{\mu}_{\tc{b}}(t)\Big].
    \end{align}

    We are interested in retrocausal contributions to the state of the detector A, thus, we find its reduced state
    \begin{equation}
        \hat{\rho}_{\tc{a}}
        = \Tr_{\tc{b}}\,\hat{\rho}_{\tc{c}} .
    \end{equation}
    A general form for the initial state of the two level systems can be expressed as:
    \begin{equation}
        \hat{\rho}_{\tc{i},0} =
        \begin{pmatrix}
        \alpha_\tc{i} & \beta_\tc{i}\\
        \beta^*_\tc{i} &1-\alpha_\tc{i}
         \end{pmatrix},
    \end{equation}
    where $\alpha_{\tc{i}}\in[0,1]$ and $ \beta_{\tc{i}}\in \mathbb{C}$.
    Using this parametrisation, up to leading order, the final state of the target detector becomes:
    \begin{equation}\label{eq:QCLeadingOrderUDW}
    \begin{aligned}
        \hat{\rho}_{\tc{a}}&=
        \hat{\rho}_{\tc{a},0}
        -\ii\lambda^2
        \!\!\int\!\!
        \dd V   
        \dd V'   
        \Delta(\mf{x},\mf{x}') 
        \Lambda_{\tc{a}}(\mf{x}) 
        \Lambda_{\tc{b}}(\mf{x}')\\
        & \times\text{Re}\left(\beta_{\tc{b}} \,e^{\ii\Omega_{\tc{b}} t}\right)
        \left[
        \hat{m}_{\tc{a}}(t'), \hat{\rho}_{\tc{a},0}
         \right].
    \end{aligned}
    \end{equation}
    The term \mbox{$\Delta(\mf{x},\mf{x}') 
    \Lambda_{\tc{a}}(\mf{x}) 
    \Lambda_{\tc{b}}(\mf{x}') \text{Re}\left(\beta_{\tc{b}} \,e^{\ii\Omega_{\tc{b}} t}\right)$} in Eq.~\eqref{eq:QCLeadingOrderUDW} determines the influence of B in the state of A. 
    We can use this prefactor to define the amount of signal that A can receive from B. 
    Notice that due to positivity of the initial state $|\Re(\beta_\tc{b}e^{\ii \Omega_\tc{b} t})|<1/\sqrt{2}$, so that the maximum signal that can be received by A is determined by the state independent term $\Delta(\mf{x},\mf{x}') 
    \Lambda_{\tc{a}}(\mf{x}) 
    \Lambda_{\tc{b}}(\mf{x}')$.
    In the same spirit as~\cite{EduCusality2015}, we define the integral of the latter term as the \textit{signalling estimator}:
    \begin{equation}\label{eq:signallingEstimator}
        C_{\tc{a}}(\Lambda_{\tc{a}}, \Lambda_{\tc{b}})= \mf{\Delta}(\Lambda_{\tc{a}}, \Lambda_{\tc{b}}) .
    \end{equation}
    Notice that the quantity above is an upper bound to the leading order change in the state of A:
    \begin{align}
        ||\hat{\rho}_\tc{a}^{(2)}|| \leq &\lambda^2 |C_\tc{a}(\Lambda_\tc{a},\Lambda_\tc{b})| \,\text{sup}_t||\hat{m}_\tc{a}(t)\hat{\rho}_{\tc{a},0} - \hat{\rho}_{\tc{a},0}\hat{m}_\tc{a}(t)||\nonumber\\
        \leq &2\lambda^2 |C_\tc{a}(\Lambda_\tc{a},\Lambda_\tc{b})|. 
    \end{align}

    Using the split $\Lambda_\tc{b} = \Lambda_\tc{b}^{(c)} + \Lambda_\tc{b}^{(r)}$ (Eq.~\eqref{eq:SplitSupport}), the signalling estimator also splits into two parts:
     \begin{equation}\label{eq:signallingEstimatorSplit}
        C_{\tc{a}}(\Lambda_{\tc{a}}, \Lambda_{\tc{b}}) = 
        C_{\tc{a}}(\Lambda_{\tc{a}}, \Lambda_{\tc{b}}^{(c)})
        +
        C_{\tc{a}}(\Lambda_{\tc{a}}, \Lambda_{\tc{b}}^{(r)}).
    \end{equation}
    The first term quantifies the signalling from the interaction subregion $\mathcal{R}_\tc{b}^{(c)}$ to A, which cannot signal retrocausally; we call that the causal part of the estimator. 
    The second term corresponds to the region that can signal retrocausally to A, $\mathcal{R}_\tc{b}^{(r)}$,(by advanced propagation); we call this the retrocausal part of the estimator. 

    In the following, we will evaluate the relative strength of these two different contributions to signalling as a function of the total duration of the interactions.

    \subsubsection{(3+1) Dimensions}
    The retarded Green's function for the (3+1)-dimensional massless scalar field (corresponding to the equation of motion~\eqref{eq:EoMScalar}), reads:
    \begin{equation}\label{eq:George}
        G_{R}(\mf{x}, \mf{x}')= \frac{1}{4 \pi |\bm{x}-\bm{x}'|}\delta(t'-t+|\bm{x}-\bm{x}'|), \quad \mathsf{x} = (\bm{x}, t).
    \end{equation}
    The  propagator \eqref{eq:George} vanishes for non null-separated events, thus, the setup of Fig.~\ref{fig:setup}  is a worst case scenario for retrocausal signalling given that if B has temporal support for time greater than $T+2L$, the additional region will not contribute to the interaction with A. 

    Let us now quantify the retrocausal contribution coming from the region that signals to A retrocausally, 
    which is of duration $2L$.
    For simplicity and since we are interested in the qualitative behaviour of the signal, we choose pointlike spatial smearing functions:
    \begin{equation}\label{eq:SpatialDeltaSmearing3}
        F_{\tc{i}}(\bm x) =\delta^{(\tc{d})}(\bm x-\bm x_{\tc{i}}).
    \end{equation}
    We also prescribe the switching functions as the following  window functions:
     \begin{equation}\label{eq:SwitchingFunction}
        \chi_{\tc{i}}(t)= \begin{cases}
            1\text{, if } t\in [t_{\tc{i}}^{\text{on}}, t_{\tc{i}}^{\text{off}}]\\
            0\text{, otherwise}
        \end{cases}.
    \end{equation}
    For this choice, the split \eqref{eq:signallingEstimatorSplit} is controlled exclusively by the switching functions that split as $\chi_{\tc{b}}  = \chi_{\tc{b}}^{(c)}+\chi_{\tc{b}} ^{(r)}$ as in Eq.~\eqref{eq:SplitSupport}.
    For the setup of Fig.~\ref{fig:setup}, we have that $t_{\tc{a}}^{(\text{off})} = t_{\tc{a}}^{(\text{on})}+T $ and
    $t_{\tc{b}}^{(\text{off})} = t_{\tc{b}}^{(\text{on})}+T +2L$, with $t_{\tc{a}}^{(\text{on})} = t_{\tc{b}}^{(\text{on})} +L$ and $|\bm x_\tc{b} - \bm x_\tc{a}| = L$, where $L$ is the spatial separation between A and B, which also corresponds to the time separation of the start of their interactions. 
    We can then write 
    \begin{equation}\label{eq:SplitSwitching}
        \begin{aligned}
        \chi_{\tc{b}}(t)&=
            1\text{, if } t\in [t_{\tc{b}}^{\text{on}}, t_{\tc{b}}^{\text{on}}+T+2L]\\
            \chi_{\tc{b}}^{(c)}(t)&=
            1\text{, if } t\in [t_{\tc{b}}^{\text{on}}, t_{\tc{b}}^{\text{on}}+T]\\\chi_{\tc{b}}^{(r)}(t)&=
            1\text{, if } t\in [t_{\tc{b}}^{\text{on}}+T, t_{\tc{b}}^{\text{on}}+T+2L],
    \end{aligned}
    \end{equation}
    with the functions vanishing outside of those intervals.
    
\begin{figure}[h]
        \centering
        \includegraphics[width=8.8cm]{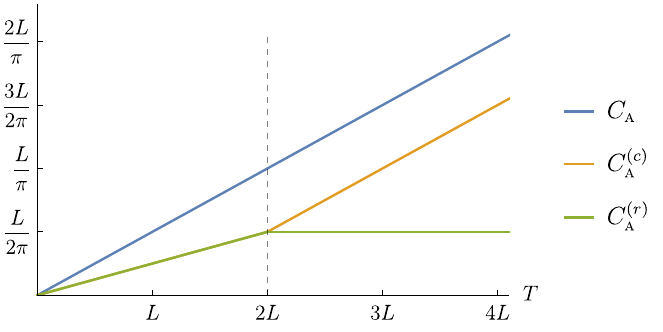}
        \caption{The dependence of the different estimators $C_{\tc{a}}$, $C_{\tc{a}}^{(c)}$, and $C_{\tc{a}}^{(r)}$ on the total time of interaction $T$ for the setup of Fig.~\ref{fig:setup} in (3+1) spacetime dimensions.}
        \label{fig:signallingScalings}
    \end{figure}

    For simplicity, we denote $C_{\tc{a}}(\Lambda_{\tc{a}}, \Lambda_{\tc{b}}) = C_{\tc{a}}$, $C_{\tc{a}}(\Lambda_{\tc{a}}, \Lambda_{\tc{b}}^{(c)}) = C_{\tc{a}}^{(c)}$, and $C_{\tc{a}}(\Lambda_{\tc{a}}, \Lambda_{\tc{b}}^{(r)}) = C_{\tc{a}}^{(r)}$. 
    The contributions to the signalling estimator defined in Eq.\eqref{eq:signallingEstimatorSplit} can be explicitly computed, yielding
    \begin{align}
        &C_{\tc{a}} =  \frac{T}{2 \pi L}\label{eq:3dFull}\\
        &C_{\tc{a}}^{(c)} = \frac{T+(-2 L+T) \Theta (T-2L)}{4 \pi L}\label{eq:3dCausal}\\
        &C_{\tc{a}}^{(r)} = \frac{T+(2 L-T) \Theta(T-2L)}{4 \pi L}\label{eq:3dRetrocausal}.
    \end{align}
    The time dependence of the different estimators is shown in Fig.~\ref{fig:signallingScalings}.
    Notice that for $T>2L$, the retrocausal contribution to the signalling estimator is constant and equal to \mbox{$C_{\tc{a}}^{(r)}= 1/(2\pi)$}.

    To compare different contributions we compute the ratios of the retrocausal signal with respect to the causal and total signal: 
    \begin{align}\label{eq:Ratios3d}
        &\frac{C_{\tc{a}}^{(r)} }{C_{\tc{a}}^{(c)} } = \frac{L}{T-L}, \text{ for } T>2 L,\\
        &\frac{C_{\tc{a}}^{(r)} }{C_{\tc{a}} } = \frac{L}{T}, \text{ for } T>2 L.
    \end{align}
    Notice that the relative impact of the retrocausal contribution decreases with the duration of the interaction, and vanishes in the limit of $T\to\infty$. Also notice that for $T<2L$ the causal and retrocausal signals contribute equally.


    Now, one might ask what would be a sufficient interaction time such that the retrocausal contribution is negligible. 
    Given a tolerance parameter, $\epsilon>0$, for the relative impact of the retrocausal signalling, $\frac{C_{\tc{a}}^{(r)} }{C_{\tc{a}} }$, the  duration of the interaction should scale as $1/\epsilon$:
    \begin{equation}
        \frac{C_{\tc{a}}^{(r)} }{C_{\tc{a}} }<\epsilon \quad \Rightarrow\quad T>\frac{L}{\epsilon}.
    \end{equation}
    That is, the retrocausal signalling predicted by the model can be neglected if the interaction time is larger than $L/\epsilon$. 
    Thus, for interactions of systems that are close and interact for sufficient time, the retrocausal contribution can be neglected.  This is consistent with what has been observed for the qc-model being able to approximate QFT mediated interactions in the limit of very long interaction times~\cite{quantClass}.

    \subsubsection{(1+1) Dimensions}

    We will now analyze the qc-model in an explicit setup in (1+1)-dimensional Minkowski spacetime. 
    The retarded propagator corresponding to the wave equation~\eqref{eq:EoMScalar} for the (1+1)-dimensional massless scalar field is given by: 
    \begin{equation}\label{eq:GR1d}
         G_{R}(\mf{x},\mf{x}') = \frac{1}{2}\Theta(t-t'-|x-x'|),
    \end{equation}    
    which is non-vanishing both for null and future time-like separation. 

\begin{figure}[h]
        \centering
        \includegraphics[width=0.5\linewidth]{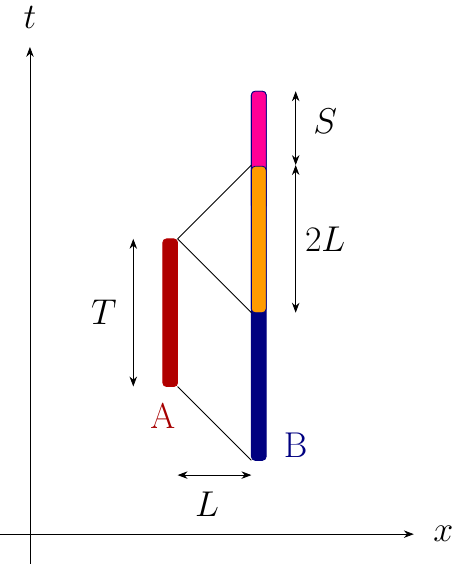}
       \caption{Spacetime diagram of the interaction regions of A and B separated by a distance $L$ when A has support contained in the future lightcone of B. The time duration of the interaction of systems A is $T$ and the time duration of B is $T+2L+S$. The region in blue signals causally to A and the orange and magenta regions are associated to retrocausal signalling from B to A.}
        \label{fig:1d}
    \end{figure}

    Given that the massless scalar field does not only propagate along light-like curves, the setup of Fig.~\ref{fig:setup} is not a worst case scenario for retrocausality. 
    Indeed, if the field propagates along time-like curves, the longer the interaction of system B lasts for, the more it can retrocausally signal to A. 
    We then consider the setup shown in Fig.~\ref{fig:1d} for our analysis in (1+1) dimensions.  
    In this setup, we will analyze the impact of the time interval of duration $2L+S$ on the final state of the system A (in contrast to the (3+1)-dimensional case where only the interval of length $2L$ needs to be taken into account).


    We repeat the analysis that we had for the (3+1)-dimensional case for the setup of Fig.~\ref{fig:1d}. 
    We also choose pointlike spatial smearing functions
    \begin{equation}\label{eq:SpatialDeltaSmearing1}
        F_{\tc{i}}(x) =\delta(x-x_{\tc{i}}),
    \end{equation}
    and we consider the same form for the switching functions of A and B as in Eq.~\eqref{eq:SwitchingFunction}.
    For the setup of Fig.~\ref{fig:1d}, we have $t_{\tc{a}}^{(\text{off})} = t_{\tc{a}}^{(\text{on})}+T $ and
    $t_{\tc{b}}^{(\text{off})} = t_{\tc{b}}^{(\text{on})}+T +2L+S$, with $t_{\tc{a}}^{(\text{on})} = t_{\tc{b}}^{(\text{on})} +L$ and $|x_\tc{b} - x_\tc{a}| = L$, where $L$ is the spatial separation between A and B.
    Then, the split of the spacetime smearings of Eq.~\eqref{eq:SplitSupport} is again controlled by the switching function of B, which can in turn be partition as    \begin{equation}\label{eq:SplitSwitching}
        \begin{aligned}
        \chi_{\tc{b}}(t)&=
            1\text{, if } t\in [t_{\tc{b}}^{\text{on}}, t_{\tc{b}}^{\text{on}}+T+2L+S],\\
            \chi_{\tc{b}}^{(c)}(t)&=
            1\text{, if } t\in [t_{\tc{b}}^{\text{on}}, t_{\tc{b}}^{\text{on}}+T],\\\chi_{\tc{b}}^{(r)}(t)&=
            1\text{, if } t\in [t_{\tc{b}}^{\text{on}}+T, t_{\tc{b}}^{\text{on}}+T+2L+S].
    \end{aligned}
    \end{equation}
    
    The signalling estimator of Eq.~\eqref{eq:signallingEstimator} and the contributions to the split in Eq.~\eqref{eq:signallingEstimatorSplit} will now depend on the times $T$ and $S$ and the separation $L$ and read
    \begin{align}
         &C_{\tc{a}} =  \frac{T(T+S)}{2}, \label{eq:1dFullSignal}\\
         &C_{\tc{a}}^{(c)} = \frac{T^{2}}{4}+\frac{(-2 L+T)^2}{4} \,\Theta(T-2 L)\label{eq:1dCausalSignal},\\
        &C_{\tc{a}}^{(r)} =\frac{T(T+2S)}{4}-\frac{(-2 L+T)^2}{4} \Theta(T-2 L)\label{eq:1dRetroSignal}.
    \end{align}
    The ratios of the retrocausal signalling estimators with respect to the causal and total signal (analogously to Eq.~\eqref{eq:Ratios3d}) are then    
    \begin{align}
        &\frac{C_{\tc{a}}^{(r)} }{C_{\tc{a}}^{(c)} } = \frac{T(2L+S)-2L^2}{ L^2 +(L-T)^2}, \text{ for } T>2 L,\\
        &\frac{C_{\tc{a}}^{(r)} }{C_{\tc{a}} } = \frac{T(2L+S)-2L^2}{T(T+S)}, \text{ for } T>2 L.
    \end{align}
    As expected, the signalling estimator in this case also depends on the time parameter $S$. 
    
    In the limit of large $S$ with constant $T$, the retrocausal contribution to the signal is dominant: 
    \begin{equation}\label{eq:AsumptRatio1d}
        \lim_{S \to \infty}\frac{C_{\tc{a}}^{(r)} }{C_{\tc{a}}^{(c)} } 
        = \infty
        \quad \text{and}\quad
       \lim_{S \to \infty} \frac{C_{\tc{a}}^{(r)} }{C_{\tc{a}} }
        = 1.
    \end{equation}
    As can be readily seen by Eq.~\eqref{eq:AsumptRatio1d}, there is no upper bound to the retrocausal signal in the (1+1)-dimensional case.
    We conclude that for (1+1)-dimensional qc-interactions the retrocausal signal can be dominant, given that B is allowed to interact for greater times than A.  

    On the other hand, the ratios of the signalling estimators for large $T$ behave as
    \begin{align}
    &\frac{C_{\tc{a}}^{(r)} }{C_{\tc{a}}^{(c)} } , \,\frac{C_{\tc{a}}^{(r)} }{C_{\tc{a}} }
        =\frac{2L+S}{T}+\mathcal{O}(T^{-2}).
    \end{align}
    Notice that if $S$ and $L$ are kept constant the ratio $ \frac{C_{\tc{a}}^{(r)} }{C_{\tc{a}}^{(c)} }$ decays to 0 as $T\to \infty$. 
    That is, even in (1+1) dimensions, in the limit of very long time interactions, the qc-model is free from retrocausal effects when both systems interact for an equally long time.

\subsection{Degenerate quantum systems}\label{sub:Gapless}
     The results of the previous subsection were obtained by a perturbative analysis, and are valid only up to leading order in the coupling constant. 
     We will now analyse retrocausal signals in the case that the two-level systems A and B have no internal dynamics, where we can solve for the final state non-perturbatively. 
     The internal dynamics are absent when the energy gap is zero, i.e. $\Omega_{\tc{i}}=0$ in Eq.~\eqref{eq:FreeHamiltonian}.
     In this case, the sources $\hat{\mu}_{\tc{a}}$ and $\hat{\mu}_{\tc{b}}$ are constant in time, so that they commute with themselves at different times, allowing both the qc-model and the QFT model to be solved exactly \cite{Landulfo:2016,analytical}. 

     For this scenario we will only be concerned with quantifying retrocausal effects in the massless (3+1)-dimensional case. 
     We will work within the setup of Fig.~\ref{fig:setup}, which is the worst case scenario for retrocausal effects in system A. 

     The interaction Hamiltonian for the interaction via the qc-field given in Eq.~\eqref{eq:QCInteractionHamiltonian} for  time independent monopoles A and B can be written in terms of the symmetric propagator as
\begin{equation}\label{eq:QCHamiltonianNP}
     \begin{aligned}
        \hat{H}_{\tc{c}}(t)
        =\frac{\lambda^2}{2}\!\!
        &\int\!\! \dd^{\textsc{d}} \bm{x}  
        \!\!\int\!\! \dd V' 
        \Delta(\mf{x},\mf{x}') 
        \Lambda_{\tc{a}}(\mf{x}) 
        \Lambda_{\tc{b}}(\mf{x}')
        \hat{\mu}_{\tc{a}}\hat{\mu}_{\tc{b}} ,
    \end{aligned}
    \end{equation}
    where $\hat{\mu}_\tc{i} $ are the time independent monopole moments of the systems A and B and satisfy $\hat{\mu}_{\tc{i}}^2 = \openone$. 
    Since the interaction Hamiltonian commutes with itself, i.e. \mbox{$[\hat{H}_{\tc{c}}(t),\hat{H}_{\tc{c}}(t')] = 0$}, 
    the unitary time evolution operator of the qc-model becomes: 
    \begin{equation}
        \hat{U}_{\tc{c}} = \exp(-\ii \int \dd t \hat{H}_\tc{c}(t)) = \exp\left(- \frac{\ii}{2} \hat{\mu}_\tc{a}\hat{\mu}_\tc{b} \mf{\Delta}_\tc{ab}\right),
    \end{equation}
   where $\mf{\Delta}_\tc{ab} = \lambda^{2}\mf{\Delta}(\Lambda_{\tc{a}}, \Lambda_{\tc{b}})$.
   Let us, then, choose some arbitrary initial state $\hat{\rho}_{0}$ for the system of A and B and the basis $\{\ket{+_\tc{a}+_\tc{b}},\ket{+_\tc{a}-_\tc{b}},\ket{-_\tc{a}+_\tc{b}},\ket{-_\tc{a}-_\tc{b}}\}$, consisting of eigenvectors of the operators $\hat{\mu}_{\tc{i}}$. 
    The final state of the two systems, $\hat{\rho} = \hat{U}_{\tc{c}}\hat{\rho}_{0}\hat{U}_{\tc{c}}^{\dagger}$, in this basis reads:
    \begin{align}
        \hat{\rho}= \sum_{i,j,k,l\in \{\pm\}}
        e^{-  \tfrac{\ii}{2}\Delta_\tc{ab}\left(
        \mu_\tc{a}^{i}\mu_\tc{b}^{j}
        -\mu_\tc{a}^{k}\mu_\tc{b}^{l}\right)}
        \proj{ij}{ij}\hat{\rho}_{0}\proj{kl}{kl},
    \end{align}
    where $\mu_\tc{i}^{\pm} = \pm 1$ are the eigenvalues of $\hat{\mu}_\tc{i}$. In matrix form, the final state reads
     \begin{equation}
        \hat{\rho} = 
        \begin{pmatrix}
         \rho_{11} 
         & \rho_{12} e^{\ii\mf\Delta_{\tc{ab}}  } 
         & \rho_{13} e^{\ii\mf\Delta_{\tc{ab}}  } 
         &\rho_{14} \\
        \rho_{21} e^{-\ii \mf\Delta_{\tc{ab}} } 
        &\rho_{\tiny{22}} 
        & \rho_{23}   
        & \rho_{24} e^{-\ii \mf\Delta_{\tc{ab}} }\\
        \rho_{31} e^{-\ii \mf\Delta_{\tc{ab}} } 
        &  \rho_{32}
        &\rho_{33} 
        &  \rho_{34} e^{-\ii \mf\Delta_{\tc{ab}} } \\
        \rho_{41} 
        &  \rho_{42} e^{\ii\mf\Delta_{\tc{ab}}  } 
        &  \rho_{43} e^{\ii\mf\Delta_{\tc{ab}}  }  
        &  \rho_{44} \\
        \end{pmatrix}\!.
    \end{equation}
    The reduced density matrix of system A is then
    \begin{equation}
        \hat{\rho}_{\tc{a}}= 
        \begin{pmatrix}
         \rho_{11} +\rho_{\tiny{22}} 
         & \rho_{13} e^{\ii\mf\Delta_{\tc{ab}}  }\! + \!\rho_{24}e^{-\ii \mf\Delta_{\tc{ab}} } \\
         \rho_{31} e^{-\ii \mf\Delta_{\tc{ab}} }
        \!+\! \rho_{42} e^{\ii\mf\Delta_{\tc{ab}}  }
        & \rho_{33}+\rho_{44} \\
        \end{pmatrix}\!.\!
    \end{equation}
    Now, we want to quantify the influence of the spacetime region of time duration $2L$ that signals to A retrocausally in Fig.~\ref{fig:setup}.
    To determine a signalling estimator in this case we find the change in the state of A after the interaction:

     \begin{align}
        \delta(\hat{\rho}_{\tc{a}})= 
        \Tr_{\tc{b}}\hat{\rho} - \Tr_{\tc{b}}\hat{\rho}_{0}& = 
        \left(e^{\ii\mf\Delta_{\tc{ab}}  } - 1\right) \begin{pmatrix}
            0&\rho_{13}\\
            \rho_{42}&0
        \end{pmatrix}\nonumber\\&
        + \left(e^{-\ii\mf\Delta_{\tc{ab}}  } - 1\right) 
        \begin{pmatrix}
            0&\rho_{24}\\
            \rho_{31}&0
        \end{pmatrix}.
    \end{align}
    The change of the state of A depends on the factors 
     \begin{align}\label{eq:factorsDeltaAB}
           \left(e^{\pm\ii\mf\Delta_{\tc{ab}}  } -1 \right)= 2\ii e^{\pm \frac{\ii}{2} \mf\Delta_\tc{ab}}\sin(\mf \Delta_{\tc{ab}}).
    \end{align}
    These factors depend exclusively on the symmetric propagator $\mf\Delta_{\tc{ab}}$, which is proportional to the signalling estimator at leading order in Eq.~\eqref{eq:signallingEstimator}. 
    Notice that one recovers the signalling estimator of Eq.~\eqref{eq:signallingEstimator} at leading order in $\lambda$ as $ \abs{e^{\pm\ii\mf\Delta_{\tc{ab}}  } -1 } = \lambda^{2} C_{\tc{a}} + \mathcal{O}(\lambda^4)$. 
    
    As we will see, both the norm and the complex argument of the terms in Eq.~\eqref{eq:factorsDeltaAB} are relevant for quantifying the signalling from B to A and discussing the regimes where retrocausal effects can be neglected. 
    We define the following quantities, that are indicative of the modulus and the argument of the factors $(e^{\pm \ii \mf \Delta_\tc{ab}}-1)$: 
    \begin{align}
        N_\tc{a}(\Lambda_\tc{a},\Lambda_\tc{b}) &\coloneqq \tfrac{1}{2}|e^{\ii\mf\Delta_{\tc{ab}}  } -1| = \abs{\sin\!\left( \tfrac{\mf\Delta_{\tc{ab}}}{2}\right)}, \label{eq:NormSinalEst}\\
        \theta_\tc{a}(\Lambda_\tc{a},\Lambda_\tc{b}) &\coloneqq \text{Arg}(-(e^{\ii\mf\Delta_{\tc{ab}}}-1)^2)=  \mf\Delta_{\tc{ab}},\label{eq:ArgSignalEst}
    \end{align}
    which we denote as $N_\tc{a}$ and $\theta_\tc{a}$, omitting the dependence on $\Lambda_{\tc{a}}, \Lambda_{\tc{b}}$. 
    Notice that the argument estimator $\theta_\tc{a}$ is proportional to the signalling estimator we defined in the perturbative case in Eq.~\eqref{eq:signallingEstimator}, i.e. $\theta_{\tc{a}} = \lambda^2C_{\tc{a}} $. 

    As before, we use pointlike spatial smearing functions as in Eq.~\eqref{eq:SpatialDeltaSmearing3} and window switching functions as in Eq.~\eqref{eq:SwitchingFunction}.
    The switching function of B splits as in Eq.~\eqref{eq:SplitSwitching}, for which we have that the symmetric propagator also splits as $\Delta_{\tc{ab}} =  \Delta_{\tc{ab}}^{(c)}+\Delta_{\tc{ab}}^{(r)}$. 
    For $T>2L$ the splitting reads
    \begin{align}
        \Delta_{\tc{ab}}^{(c)} =\frac{\lambda^2T}{2\pi L}, \quad
        \Delta_{\tc{ab}}^{(r)} =  \frac{\lambda^2}{2\pi}.
    \end{align}
    Evidently, the argument $\theta_{\tc{a}} = \theta_\tc{a}^{(c)} + \theta_\tc{a}^{(r)}$ splits identically so that for $T>2L$
    \begin{align}\label{eq:Arg}
         \theta_\tc{a}^{(c)} =\frac{\lambda^2T}{2\pi L}, \quad  \theta_\tc{a}^{(r)} =\frac{\lambda^2}{2\pi}.
    \end{align}
    The dependence of $N_\tc{a}$ in the retrocausal region, on the other hand, is non-linear, so it does not split as a sum $N_\tc{a} \neq N_\tc{a}^{(c)} + N_\tc{a}^{(r)}$.  
    To estimate retrocausal effects using $N_\tc{a}$ we can instead compare $N_\tc{a}$ and $N_\tc{a}^{(c)} = \abs{\sin\!\left( \tfrac{\mf\Delta_{\tc{ab}}^{(c)}}{2}\right)}$, defined when considering only the causal part of the interaction. For $T>2L$ we find
    \begin{align}\label{eq:NormSplit}
        N_\tc{a}&=\abs{\sin\!\left( \tfrac{\lambda^2T}{4\pi L} + \tfrac{\lambda^2}{4\pi }\right)}\nonumber ,\\
        N_\tc{a}^{(c)} &= \left|\sin(\tfrac{\lambda^2T}{4 \pi L})\right|.
    \end{align}
        
    Now that we have analyzed the causal and retrocausal contributions to the estimators we discuss their behaviour with respect to interaction time $T$ and the coupling $\lambda$.     
    To study the retrocausal contribution to $N_{\tc{a}}$ we subtract $N_\tc{a}^{(c)}$ to obtain the following upper bound for the difference:
   \begin{align}
        \abs{N_\tc{a} - N_{\tc{a}}^{(c)}} <\epsilon.
    \end{align}
    Now notice that the difference of the two is bounded by the expression
    \begin{align}
        \abs{N_\tc{a} - N_{\tc{a}}^{(c)}}  = \abs{\abs{\sin\!\left( \tfrac{\lambda^2T}{4\pi L} + \tfrac{\lambda^2}{4\pi }\right)} - \abs{\sin\!\left( \tfrac{\lambda^2T}{4\pi L}\right)}}<\epsilon.
    \end{align}
    For small $\epsilon$, the inequality above is satisfied when
    \begin{equation}\label{eq:CouplingConstraint}
        \frac{\lambda^2}{4\pi} \lesssim \epsilon.
    \end{equation}

    We conclude from the above that for the retrocausal contribution to not have significant effect, sufficiently large interaction times and small enough couplings are required\footnote{Notice that small $\lambda$ does not imply that we are in the perturbative regime, as the small terms in the perturbative regime are of the order of $\lambda^2T/L$. The non-perturbative regime then allows one to consistently take the limit $T\to\infty$.}. 
    For the argument $\theta_{\tc{a}}$, the retrocausal contribution will always be negligible for very large times independently of the coupling.
    If the coupling is very strong, though, the phase shift of $N_{\tc{a}}$ by $\frac{\lambda^2}{4\pi}$ will be significant\footnote{Notice that, rigorously, the model will also provide negligible retrocausal effects for strong coupling as long as $|\frac{\lambda^2}{4\pi} - 2 n \pi|<\epsilon$ for $n\in\mathbb{N}$, and not just small $\lambda$. Intuitively, this is because for these specific values of $\lambda$ the retrocausal interaction region $\Lambda_\tc{b}^{(r)}$ very approximately acts as the identity on system A and can yield a negligible contribution to the time evolution when compared to $\theta_\textsc{a}^{(c)}$.}.

    For small enough $\lambda$ (that is, $\lambda^2 \ll L/T$), as far as the argument $\theta_{\tc{a}}$ is concerned, as $T$ increases the constant retrocausal contribution becomes negligible compared to the causal one in Eq.~\eqref{eq:Arg}, as the ratio $\frac{\theta_{\tc{a}}^{(r)}}{\theta_{\tc{a}}^{(c)}}$ decays as $\sim\frac{1}{T}$. 
    Given a tolerance $\epsilon>0$ for the retrocausal contribution, we find the condition 
    \begin{equation}
        \frac{\theta_\tc{a}^{(r)}}{\theta_\tc{a}^{(c)}} < \epsilon \Rightarrow  T> \frac{L}{\epsilon},
    \end{equation}  
    which is the same condition that we found for the signalling estimator $C_{\tc{a}}$ in the perturbative case.

    It is important to appreciate that the retrocausal contribution to the causality estimators $\theta_\tc{a}$ and $N_\tc{a}$ appears as a phase shift that is entirely determined by the coupling constant $\lambda$. 
    Being a phase, the retrocausal contribution does not become negligible for non-perturbative $\lambda$ in the limit of long interaction times, since it simply shifts the periodic function $N_{\tc{a}}^{(c)}$ in time $T$.
    However, the phase shift can indeed yield non-measurable effects: for experimental setups where phase shifts of the order of $\frac{\lambda^2}{4\pi}$ cannot be accurately resolved the retrocausal effects would not be discernible. 
    The time resolution required to measure the phase shift can be estimated by analyzing the period of the modulus and phase of the signalling estimator.
    Reformulating $N_{\tc{a}}$ as $N_\tc{a}=\abs{\sin\! \left(\tfrac{\lambda^2}{4\pi L} (T+L)\right)}$, we see that the time shift in $N_\tc{a}$ (and also in $\theta_\tc{a}$) is $\delta T =L$ and the period $T_{p} = \tfrac{8\pi^2 L}{\lambda^2}$, so the necessary time resolution where the retrocausal effects can be discerned (and therefore the qc-model would not be a good description of the physical scenario) is the minimum of the two.

    \subsection{Applicability of the qc-model to gravity mediated entanglement experiments}

    In this subsection we are concerned with the time scales where retrocausal effects predicted by the qc-model could be observable.
    These are regimes where the qc-model cannot be used to describe a physical situation. 

    The main motivation for this analysis stems from gravity mediated entanglement (GME) experiments that aim to witness quantum degrees of freedom of the gravitational field  \cite{B,MV}. 
    In summary, the experimental proposals of~\cite{B,MV} the masses that get gravitationally entangled are separated by distances of the order of $L\sim 10^{-6}\rm m$ and interact for times of the order of $T\sim 1\rm{s}$. 
    As has been pointed out in~\cite{ourBMV}, experiments in these regimes do not rely on gravitational degrees of freedom and can instead be modelled via a qc-interaction. 
    However, as already discussed, there are regimes where the qc-model cannot be an acceptable effective model for interaction scenarios due to retrocausal effects.    
    This raises the question of whether retrocausal effects present in the qc-model can be used to argue that the observed gravity induced entanglement corresponds to a genuine quantum description of gravity~\cite{FlaminiaQuantumGravity} for the time scales in current proposals for experimental tests. 


    As we saw for the massless scalar field in (3+1)-dimensions the retrocausal contributions in the qc-model can be neglected for sufficiently long interaction times and small couplings. 
    This analysis directly generalises to massless fields of higher spins such as gravitational perturbations\footnote{For a metric perturbation $h_{\mu\nu} = \sqrt{4 \pi G} \gamma_{\mu\nu}$, a localized system of mass $m$ interacts with the effective scalar field $\phi(\mf x) = -\tfrac{1}{2}u^\mu u^\nu \gamma_{\mu\nu}(\mf x)$, where $u^\mu$ is its four-velocity and the effective coupling constant becomes $\lambda = m \sqrt{\pi G}$. The qc-description for the interaction of two masses was explicitly given in~\cite{ourBMV}, where it was shown that the interaction behaves exactly like the scalar interaction, apart from a potential redshift factor arising from the different four-velocities of the two systems (see the supplemental material of~\cite{ourBMV}).}. 
    The effective coupling constant $\lambda^2$ in the gravitational case is proportional to $G m_1 m_2$, where $m_1$ and $m_2$ are the masses of the particles used in the experiments. Typical experimental values in current proposals are are of the order of $m_1,m_2\sim 10^{-14}\text{kg}$~\cite{B,MV}. 
    Considering the currently proposed values of $L$ and $T$ for gravity mediated entanglement experiments, we find that $T/L \sim 10^{14}$ and the effective coupling constant takes the value of $\lambda^2 \sim  10^{-14}$, both of which are well within the regime where the retrocausal effects of the qc-model can be neglected or not discerned. Indeed, our analysis shows that these retrocausal effects would only be observable in comparison with the experimental observation if the available time resolution is at least of the order of $\delta T\sim L \sim 10^{-14}\rm{s}$.

    We can then conclude that although the qc-model is not fully compatible with relativistic principles (because it predicts retrocausal effects), the regimes of validity of the model are well within the bounds of current experimental proposals to witness gravity induced entanglement.
    This shows that unless our time resolution for GME experiments increases significantly, there is still a possible description for the current  proposals that does not rely on degrees of freedom for the gravitational field and is experimentally indistinguishable from a fully quantum description for the gravitational field along the lines of~\cite{ourBMV}.

\section{Conclusions} \label{sec:conclusions}

In this work we studied the restrictions imposed by demanding insignificant retrocausal effects in the predictions of the  quantum-controlled model. Qc-models are an effective description of relativistically propagating interactions between quantum systems through a mediating field, but the field has no quantum quantum degrees of freedom. 
As such, the regime of applicability of the qc-model is also indicative of the regimes where physical processes are insensitive to the quantumness of the field. This is particularly relevant in the context of gravity mediated entanglement (GME) experiments, which have been proposed as a way to witness the quantum nature of the gravitational field.

Concretely, we provided an explicit definition of retrocausality in relativistic interactions, from which we were able to separate the causal and retrocausal contributions to the predictions of qc-interactions. 
This allowed us to explicitly quantify retrocausal signalling in the qc-model. 
We found that, while retrocausal effects are intrinsic to qc-models, they nonetheless become negligible compared to the causal signals in the limit where systems maintain causal contact for long times relative to their spatial separation. 
This limit of long interactions is also precisely the regime where the qc-model was shown to be a good approximation to quantum field theory~\cite{quantClass}.

For interactions mediated by massless fields in (3+1) dimensional Minkowski spacetime (such as linearized gravitational perturbations), we concluded that retrocausal effects in qc-models are upper bounded by the square of the interaction strength. We applied our results to the specific reference parameters of the proposed GME experimental setups of~\cite{B}. We found that to rule out a possible qc-description due to retrocausal effects one needs to be able to distinguish relative variations of the quantum state of the sources of the order of $10^{-14}$, or a time resolution of the order of the light-crossing time between the systems. Overall, our results reinforce the findings of~\cite{quantClass} that processes that involve weak couplings and long interaction times cannot resolve the quantum degrees of freedom of a field theory.

\acknowledgements

TRP acknowledges support from the Natural Sciences and Engineering Research Council of Canada (NSERC) via the Vanier Canada Graduate Scholarship. EMM acknowledges support through the Discovery Grant Program of the Natural Sciences and Engineering Research Council of Canada (NSERC). EMM also acknowledges support of his Ontario Early Researcher award. Research at Perimeter Institute is supported in part by the Government of Canada through the Department of Innovation, Science and Industry Canada and by the Province of Ontario through the Ministry of Colleges and Universities. Perimeter Institute and the University of Waterloo are situated on the Haldimand Tract, land that was promised to the Haudenosaunee of the Six Nations of the Grand River, and is within the territory of the Neutral, Anishinaabe, and Haudenosaunee people.
    
\appendix



\section{The Hamiltonian for the classical scalar field coupled to classical sources}\label{app:HamiltonianProof}

In this Appendix we show that the interaction of two systems through a mediating field can be modelled by the interaction Hamiltonian of Eq.~\eqref{eq:HamiltonianClassical} under the assumption that the field is entirely determined by sources that are slowly varying in time when we neglect self-interactions. We will explicitly show this for a scalar field $\phi(\mf x)$ of mass $m$, but our results carry on to more general interactions, such as masses interacting via the gravitational field. 

We consider two systems A and B coupled to the scalar field via the scalar current densities $j_{\tc{a}}(\mf x)$, $j_{\tc{b}}(\mf x)$.
The action that describes the dynamics of the sources and the field, as well as the interactions between them, is analogous to Eq.~\eqref{eq:LagrangianClassical}
\begin{equation}\label{eq:app:action}
    S = \!\int \!\dd V \!\left(\L_\textsc{a}\!+\!\L_\textsc{b}\! + \!\L_\phi \!-j(\mf x) \phi(\mf x) \right),
\end{equation}
where $j(\mf x)=j_{\tc{a}}(\mf x)+j_{\tc{b}}(\mf x)$.
The corresponding Hamiltonian density is
\begin{equation}\label{eq:app:HamiltDensity}
    \mathcal{H} = \mathcal{H}_\textsc{a}\!+\!\mathcal{H}_\textsc{b}\! + \!\mathcal{H}_\phi\!+j(\mf x) \phi(\mf x), 
\end{equation}
where $\mathcal{H}_\textsc{a}$, $\mathcal{H}_\textsc{b}$ are the free Hamiltonian densities for systems A and B and $\!\mathcal{H}_\phi$ is the free Hamiltonian density for the scalar field, explicitly given by
\begin{equation}
    \mathcal{H}_\phi = (\partial_t\phi)^2 + \frac{1}{2} \left(\partial_\mu \phi \partial^\mu\phi + m^2\phi^2\right).
\end{equation}
The Hamiltonian associated to an inertial coordinate system $(t,\bm x)$ is obtained by integrating Eq.~\eqref{eq:app:HamiltDensity} over the spatial variables:
\begin{equation}\label{eq:app:HamSplitbforeAss}
    H = H_\tc{a} + H_\tc{b} + H_\phi + \int \dd^n \bm x j(\mf x) \phi(\mf x).
\end{equation}
Our goal in this appendix is to show that the Hamiltonian $H$ can be approximated by
\begin{align}
    H \approx \,& H_\tc{a} + H_\tc{b} \label{eq:HgoalApp}\\
    &+ \frac{1}{2} \int \!\dd^n
\bm x \!\int\!\! \dd V' G_R(\mf x, \mf x') (j_\tc{a}(\mf x)j_\tc{b}(\mf x') + j_\tc{a}(\mf x')j_\tc{b}(\mf x)),\nonumber
\end{align}
the analogue of Eq.~\eqref{eq:HamiltonianClassical} for a scalar field.

We start by rewriting the free Hamiltonian density of the scalar field as
\begin{align}
    \mathcal{H}_\phi &=  (\partial_t\phi)^2  +\frac{1}{2}\partial_{\mu}\left(\phi\partial^{\mu}\phi\right) - \frac{1}{2}\phi\partial_{\mu}\partial^{\mu}\phi +\frac{m^2}{2}\phi^2\\
    &=  (\partial_t\phi)^2  +\frac{1}{2}\partial_{\mu}\left(\phi\partial^{\mu}\phi\right) -\frac{1}{2}\phi (\partial_\mu \partial^\mu - m^2)\phi.
\end{align}
Notice that $\mathcal{P}[\phi]\coloneqq(\partial_{\mu}\partial^{\mu} - m^2)\phi$ is the Klein-Gordon differential operator that defines the equation of motion for the field $\phi(\mf x)$. In particular, in the presence of the source $j(\mf x)$, the equation of motion for the field becomes$\mathcal{P}[\phi] = j(\mf x)$, so that 
\begin{align}
    \mathcal{H}_\phi  
    =  (\partial_t\phi)^2  +\frac{1}{2}\partial_{\mu}\left(\phi\partial^{\mu}\phi\right) -\frac{1}{2}\phi(\mf x) j(\mf x).
\end{align}

We can now integrate $\mathcal{H}_\phi$ over a $t = \text{const.}$ spatial slice to obtain the Hamiltonian
\begin{align}\label{eq:theoneabove}
    H_\phi &= \int \dd^n \bm x \left((\partial_t\phi)^2  - \frac{1}{2}\partial_{t}\left(\phi\partial_t\phi\right)- \frac{1}{2} \phi(\mf x)  j(\mf x)\right)\\ &= \frac{1}{2}\int \dd^n \bm x \left((\partial_t\phi)^2  - \phi \partial_t^2 \phi\right) - \frac{1}{2} \int \dd^n \bm x \phi(\mf x) j(\mf x),
\end{align}
where the terms $\partial_i(\phi \partial^i \phi)$ are total spatial derivatives, which reduce to boundary terms that vanish under the assumption that the field decays sufficiently fast at spatial infinity. Using that the field is entirely determined by the sources, this assumption is equivalent to the statement that the sources are localized in space. 

Using Eq.~\eqref{eq:theoneabove} in Eq.~\eqref{eq:app:HamSplitbforeAss} we can recast it as
\begin{align}
    H = H_\tc{a} + H_\tc{b} &+ \frac{1}{2} \int \dd^n\bm x j(\mf x) \phi(\mf x) \\
    &\:\:\:\:+ \frac{1}{2}\int \dd^n \bm x \left((\partial_t\phi)^2  - \phi \partial_t^2 \phi\right).\nonumber
\end{align}
For convenience, we define
\begin{align}
    H_{\text{int}} &\coloneqq \frac{1}{2} \int \dd^n\bm x j(\mf x) \phi(\mf x),\\
   H_\text{diff}&\coloneqq \frac{1}{2}\int \dd^n \bm x \left((\partial_t\phi)^2  - \phi \partial_t^2 \phi\right).
\end{align}
As we will see, the Hamiltonian $H_\text{int}$ will correspond to the retarded propagated interaction between the sources and $H_\text{diff}$ will be negligible under the assumption that the sources are slowly varying in time.

Let us first rewrite $H_\text{int}$ in terms of retarded Green's functions. Given that the field satisfies the equation of motion $\mathcal{P}[\phi] = j$ and is entirely sourced by $j(\mf x)$, it can be written in terms of the retarded Green's function of the operator $\mathcal{P}$:
\begin{equation}\label{eq:app:FieldSourced}
    \phi(\mf x) =  \int \dd V' G_R(\mf x,\mf x') j(\mf x').
\end{equation}
This allows us to write 
\begin{equation}
    H_\text{int} = - \frac{1}{2}\int \dd^n \bm x \!\!\int \!\!\dd V' j(\mf x) G_R(\mf x, \mf x')j(\mf x'). 
\end{equation}
Moreover, using $j(\mf x) = j_\tc{a}(\mf x) + j_\tc{b}(\mf x)$ and ignoring the self-interaction terms, we obtain
\begin{align}
    H_\text{int}= - \frac{1}{2}\!\!\int \!\dd^n \bm x \!\!\int \!\!\dd V'G_R(\mf x, \mf x') (j_{\tc{a}}(\mf x)j_{\tc{b}}(\mf x')+j_{\tc{a}}(\mf x')j_{\tc{b}}(\mf x)),
\end{align}
which precisely matches the interaction Hamiltonian of Eq.~\eqref{eq:HamiltonianInterClassical}. 

The final step to show that Eq.~\eqref{eq:HgoalApp} holds is to demonstrate that $H_\text{diff}\approx 0$ in the limit where the sources are slowly varying with time. First notice that $H_\text{diff}$ involves time derivatives of the field $\phi(\mf x)$. Using Eq.~\eqref{eq:app:FieldSourced}, we can write
\begin{align}
    \partial_t \phi(\mf x) &= - \int \dd V' G_R(\mf x, \mf x') \partial_{t'}j(\mf x'),\\
    \partial_t^2 \phi(\mf x) &= \int \dd V' G_R(\mf x, \mf x') \partial_{t'}^2j(\mf x'),
\end{align}
from which we conclude that $H_\text{diff}$ depends explicitly on the time derivatives of the sources.

To show that the $H_\text{diff}$ is negligible if the sources vary slowly, we explicitly write the field $\phi(t,\bm x)$ sourced by $j(t,\bm x)$ in terms of a profile function $\Phi(s,\bm x)$ and a time parameter $T$ that controls the time variation of the field:
\begin{align}
    \phi(t,\bm x) = \Phi(t/T,\bm x).
\end{align}
Given that $\Phi(s,\bm x)$ is a profile function, its integrals in space, as well as the integrals of its derivatives with respect to the dimensionless parameter $s$ are constant and independent of the time parameter $T$.
We can rewrite $H_\text{diff}$ in terms of the profile $\Phi(s,\bm x)$:
\begin{equation}
    H_{\text{diff}} = \frac{1}{2T^2}\int \dd^n \bm x \left(\left(\partial_s\Phi)^2  - \Phi \partial_s^2 \Phi\right)\right|_{s = \tfrac{t}{T}}.
\end{equation}
Given that $\left( (\partial_{s}\Phi)^{2} - \Phi \partial_{s}^{2} \Phi\right)$ is bounded, for large time scales $T$, the term $\mathcal{H}_{\text{diff}}$ becomes negligible. The limit of large $T$ entails to an adiabatic approximation, where the fields (thus the current densities) only vary significantly over times of the same scales as $T$.

\bibliography{bibliography}

\end{document}